\documentclass[sigconf, nonacm]{acmart}

\usepackage[noend]{algpseudocode}
\usepackage{algorithm}
\usepackage{bbm}
\usepackage{graphicx}
\usepackage{listings}
\usepackage[compress]{cleveref}
\usepackage{mathrsfs}
\usepackage{bm}
\usepackage{balance}
\usepackage{outlines}
\usepackage{svg}
\usepackage{grffile}
\usepackage[caption=false]{subfig}
\usepackage{xspace}
\usepackage{enumitem}
\usepackage{xcolor}
\usepackage{array}
\usepackage{nccmath}
\usepackage{multirow}
\usepackage{pifont}
\usepackage{soul}
\usepackage{caption}
\usepackage{upgreek}
\usepackage{bbm}
\usepackage{dsfont}
\usepackage{microtype}
\usepackage{amsfonts}
\usepackage{amsmath}
\usepackage{accents}
\usepackage{tabularx}
\usepackage[group-separator={,}]{siunitx}
\usepackage[bottom]{footmisc}
\usepackage{setspace}
\usepackage{etoolbox}

\makeatletter
\preto{\@verbatim}{\topsep=0pt \partopsep=0pt }
\makeatother

\newcommand{\ubar}[1]{\underaccent{\bar}{#1}}

\newcommand{\lowoverline}[1]{%
  \overline{\smash{#1}\vphantom{x}}\vphantom{#1}%
}

\usepackage{xpatch}
\xpatchcmd{\NCC@ignorepar}{%
\abovedisplayskip\abovedisplayshortskip}
{%
\abovedisplayskip\abovedisplayshortskip%
\belowdisplayskip\belowdisplayshortskip}
{}{}

\graphicspath{./figures/}

\crefformat{section}{\S#2#1#3} %
\crefformat{subsection}{\S#2#1#3}
\crefformat{subsubsection}{\S#2#1#3}
\crefname{section}{Section}{Sections}
\Crefname{section}{Section}{Sections}
\crefname{figure}{Figure}{Figures}
\Crefname{figure}{Figure}{Figures}
\crefname{subfigure}{Figure}{Figures}
\Crefname{subfigure}{Figure}{Figures}
\crefname{algorithm}{Algorithm}{Algorithms}
\Crefname{algorithm}{Algorithm}{Algorithms}
\crefname{equation}{Equation}{Equation}
\Crefname{equation}{Equation}{Equation}
\crefname{lemma}{Lemma}{Lemma}
\Crefname{lemma}{Lemma}{Lemma}
\crefname{table}{Table}{Tables}
\Crefname{table}{Table}{Tables}
\crefrangelabelformat{subfigure}{#3#1#4--#5(\crefstripprefix{#1}{#2}#6}

\definecolor{emerald}{rgb}{0.31, 0.78, 0.47}

\newcommand{\revisioncolor}{black}
\newcommand{\revision}[1]{{\color{\revisioncolor} #1}}

\newcommand{\eat}[1]{}

\newcommand{\preeq}{\vspace{0mm}\begin{small}}
\newcommand{\posteq}{\vspace{0mm}\end{small}}

\newcommand{\system}{\textsc{MaskSearch}\xspace}

\usepackage{url}

\DeclareFixedFont{\ttb}{T1}{txtt}{bx}{n}{8}
\DeclareFixedFont{\ttm}{T1}{txtt}{m}{n}{8}

\usepackage{color}
\definecolor{deepblue}{rgb}{0,0,0.5}
\definecolor{deepred}{rgb}{0.6,0,0}
\definecolor{deepgreen}{rgb}{0,0.5,0}
\definecolor{purple}{rgb}{0.5,0,0.5}
\definecolor{gray}{rgb}{0.33,0.33,0.33}

\definecolor{dkgreen}{rgb}{0,0.6,0}
\definecolor{gray}{rgb}{0.5,0.5,0.5}
\definecolor{mauve}{rgb}{0.58,0,0.82}

\lstdefinelanguage{Python}{
	keywords={typeof, torch, nonzero, index_select, zeros_like, lt, masked_select, new, true, false, catch,def,val, function, return, null, catch, switch, var, shape,  while, do, else, case, break, override},
	keywordstyle=\color{blue}\bfseries,
	ndkeywords={class, export,extends, boolean, throw, implements, import, this, abstract, for, in, if},
	ndkeywordstyle=\color{dkgreen}\bfseries,
otherkeywords={+, =>,<=, ==, >,< , || , T},
	identifierstyle=\color{black},
	sensitive=false,
	comment=[l]{//},
	morecomment=[s]{/*}{*/},
	commentstyle=\color{purple}\ttfamily,
	stringstyle=\color{red}\ttfamily,
	morestring=[b]',
	morestring=[b]"
}
\begin{document}

\title{\system: Querying Image Masks at Scale}

\author{Dong He, Jieyu Zhang, Maureen Daum, Alexander Ratner, Magdalena Balazinska}

\affiliation{%
  \institution{University of Washington,\quad \{donghe, jieyuz2, mdaum, ajratner, magda\}@cs.washington.edu}
  \city{}
  \country{}
}

\begin{abstract}
\begin{sloppypar}
Machine learning tasks over image databases often generate masks that annotate image content (e.g., saliency maps, segmentation maps, depth maps) and enable a variety of applications (e.g., determine if a model is learning spurious correlations or if an image was maliciously modified to mislead a model). 
While queries that retrieve examples based on mask properties are valuable to practitioners, existing systems do not support them efficiently. 
In this paper, we formalize the problem and propose \system, a system that focuses on accelerating queries over databases of image masks while guaranteeing the correctness of query results. 
\system leverages a novel indexing technique and an efficient filter-verification query execution framework. 
Experiments with our prototype show that \system, using indexes approximately $5\%$ of the compressed data size, accelerates individual queries by up to two orders of magnitude and consistently outperforms existing methods on various multi-query workloads that simulate dataset exploration and analysis processes. 
\end{sloppypar}
\end{abstract}

\maketitle

\begin{sloppypar}

\newcommand{\maskExampleFigure}{
    \begin{figure}[t!]
        \begin{center}
            \subfloat[Segmentation mask]{\includegraphics[width=0.34\linewidth]{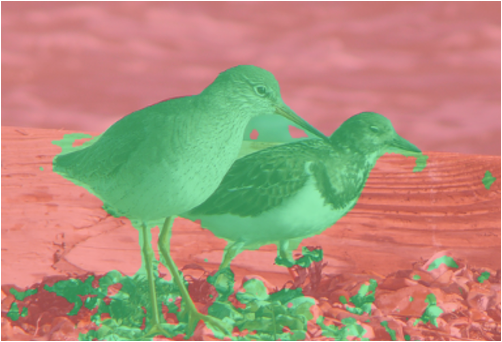}%
                \label{subfig:image-segmentation-example}}%
            \hfil
            \subfloat[Depth estimation mask]{\includegraphics[width=0.338\linewidth]{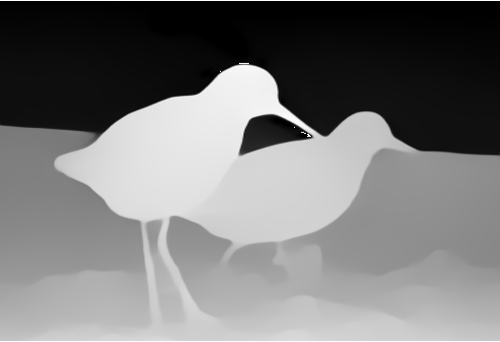}%
                \label{subfig:depth-estimation-example}}%
            \hfil
            \subfloat[Saliency map]{\includegraphics[width=0.25\linewidth]{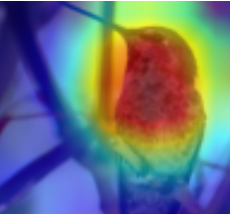}%
                \label{subfig:saliency-map-example}}%
        \end{center}
        \vspace{-1.4em}
        \caption{Examples of image masks that annotate image content for ImageNet~\cite{ILSVRC15} images produced by ML tasks.}
        \vspace{-1em}
        \label{fig:mask-example}
    \end{figure}
}

\newcommand{\introExampleFigure}{
    \begin{figure}[t!]
        \centering
        \includegraphics[width=0.85\linewidth]{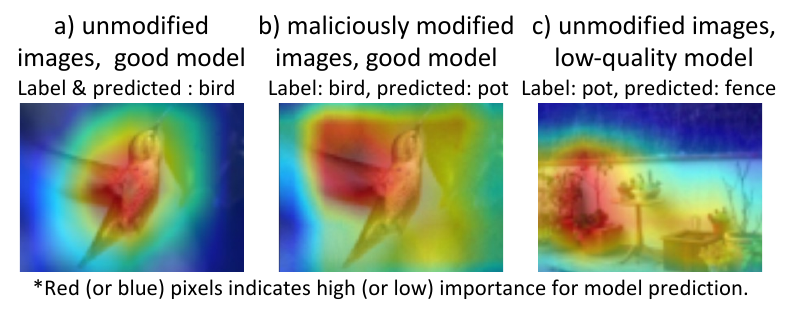}
        \vspace{-1.5em}
        \caption{
        Example image masks: ImageNet~\cite{ILSVRC15} images overlaied with saliency maps. 
        Saliency maps in columns b) and c) reveal that the models rely on irrelevant pixels to make predictions. 
        Retrieving more examples with similar mask properties helps to better investigate the model's behavior. 
        }
        \vspace{-1em}
        \label{fig:intro-example}
    \end{figure}
}

\newcommand{\maliciousAttackFigure}{
    \begin{figure}[t!]
        \centering
        \includegraphics[width=\linewidth]{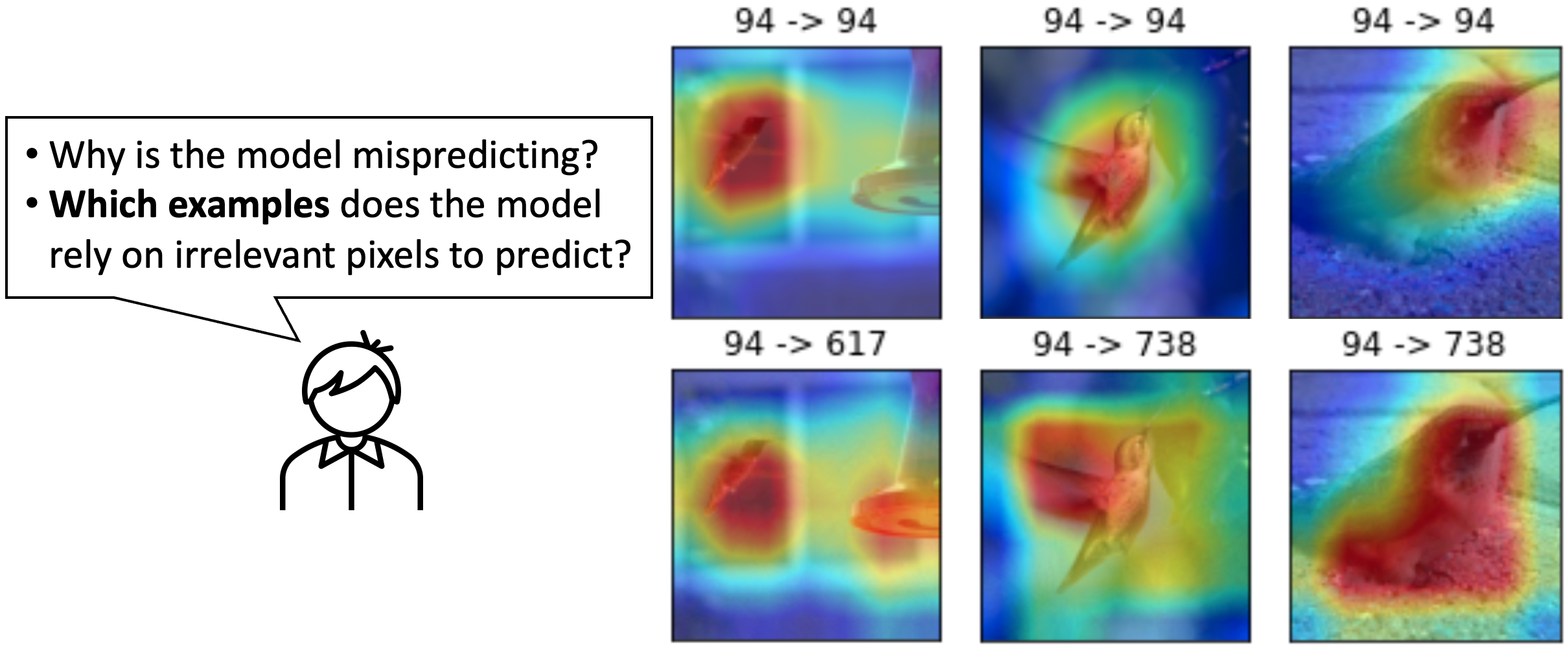}
        \vspace{-1.2em}
        \caption{.}
        \vspace{-0.53em}
        \label{fig:malicious-attack}
    \end{figure}
}

\newcommand{\queryExampleFigure}{
    \begin{figure}[t!]
        \begin{center}

            \subfloat[Example image]{\includegraphics[width=0.25\linewidth]{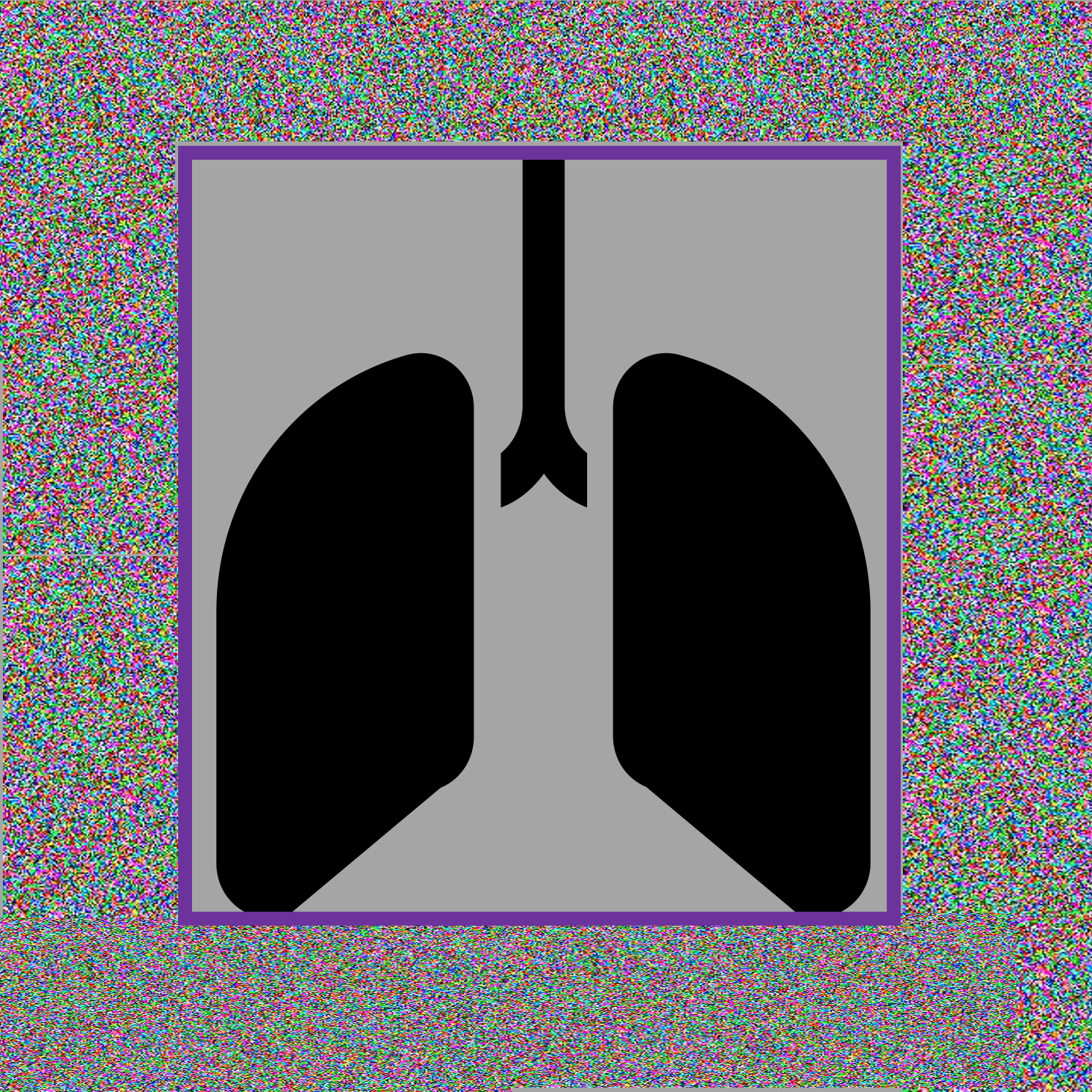}%
                \label{subfig:query-example-image}}%
            \hfil
            \subfloat[Mask]{\includegraphics[width=0.25\linewidth]{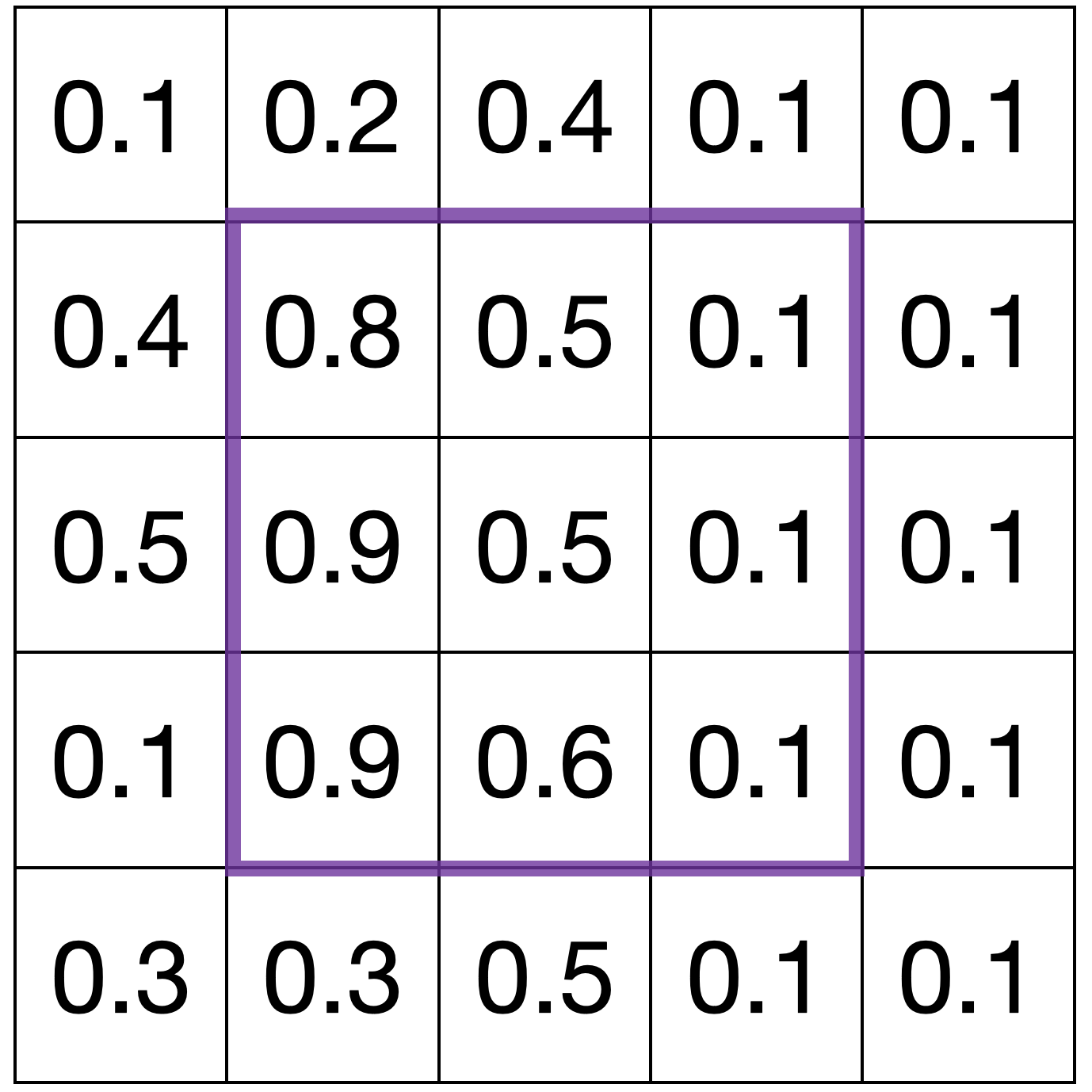}%
                \label{subfig:query-example-mask}}%
        \end{center}
        \vspace{-1.2em}
        \caption{A toy image motivated by~\cite{degrave2021ai} and its mask. The purple box is the ROI. Predicates on masks often involve counting the number of pixels in the ROI with values in a range, e.g., $\#$ pixels in the ROI with values in $(0.85, 1.0)$ is $2$.}
        \vspace{-1.0em}
        \label{fig:query-example}
    \end{figure}
}

\newcommand{\notationTable}{
    \begingroup
    \setlength{\tabcolsep}{2pt}
    \begin{table}[t!]
        \small
        \centering
        \caption{Summary of frequently used notation.}
        \vspace{-1.1em}
        \begin{tabular}{ l l }
            \toprule
            Symbol & Meaning \\
            \midrule
            $\texttt{CP}(mask, r, (lv, uv))$ & Count of pixels in region $r$ of $mask$ \\
             & with pixel values in range $(lv, uv)$ \\
            $P$ & Predicate $\texttt{CP}(mask, roi, (lv, uv)) > T$ \\
            $\theta$ & Actual value of $\texttt{CP}(mask, roi, (lv, uv))$ \\
            $\bar{\theta}$ & Upper bound of $\theta$ \\
            $\ubar{\theta}$ & Lower bound of $\theta$ \\
            $\Delta$ & Size of a pixel value bin \\
            $C(mask\_id, r)$ & Histogram of reverse cumulative pixel counts \\
            $C(mask\_id, r)[i]$ & $\texttt{CP}(mask, r, (p_{min} + i\Delta, p\_max))$ \\
            $roi$ & Region of interest specified by the user \\
            $\lowoverline{roi}$ & Smallest region \textit{available} in CHI covering $roi$ \\
            $\underline{roi}$ & Largest region \textit{available} in CHI covered by $roi$ \\
            \bottomrule
        \end{tabular}
        \normalsize
        \label{tab:notation}
    \end{table}
    \endgroup
}

\newcommand{\chiIllustrationFigure}{
    \begin{figure}
    \centering
    \includegraphics[width=\linewidth]{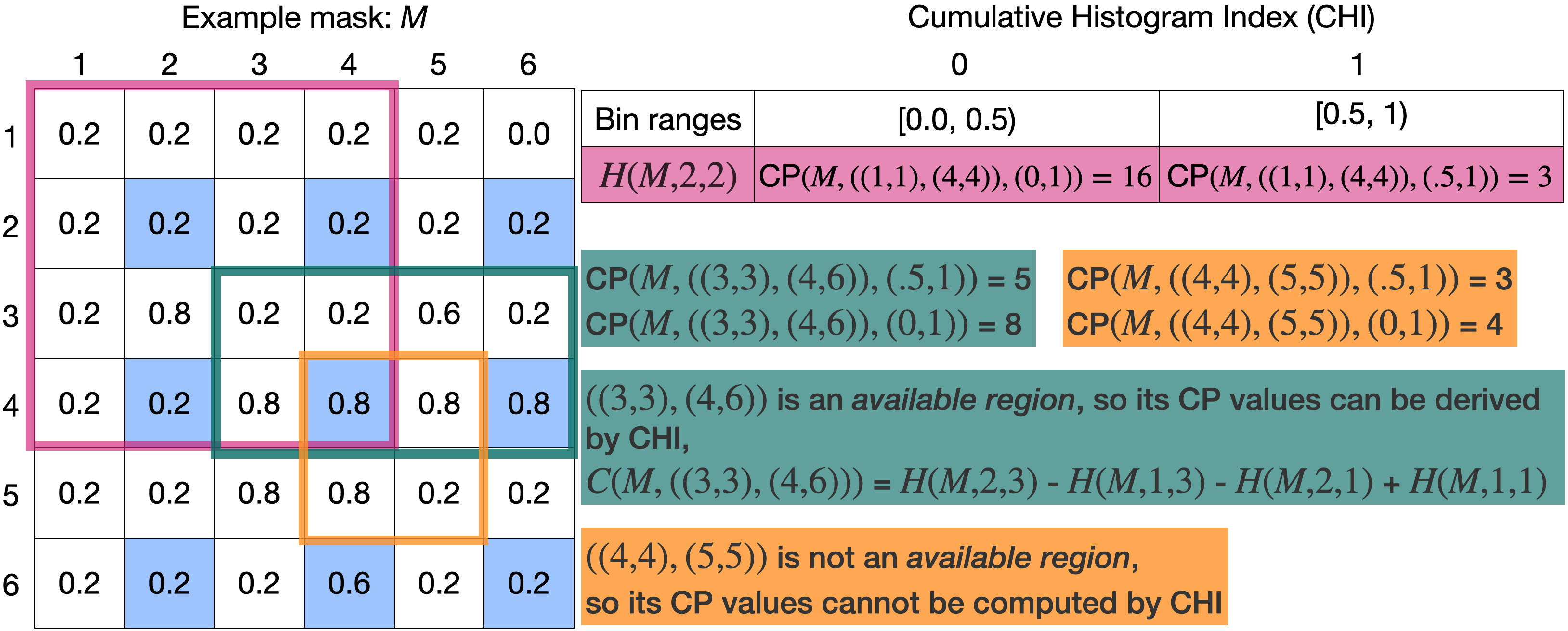}
    \vspace{-2.0em}
    \caption{An example of CHI, \texttt{CP}, \textit{available region}, and $C$.}
    \vspace{-1.0em}
    \label{fig:chi-illustration}
    \end{figure}
}

\newcommand{\upperBoundIllustrationFigure}{
    \begin{figure}
    \centering
    \includegraphics[width=0.85\linewidth]{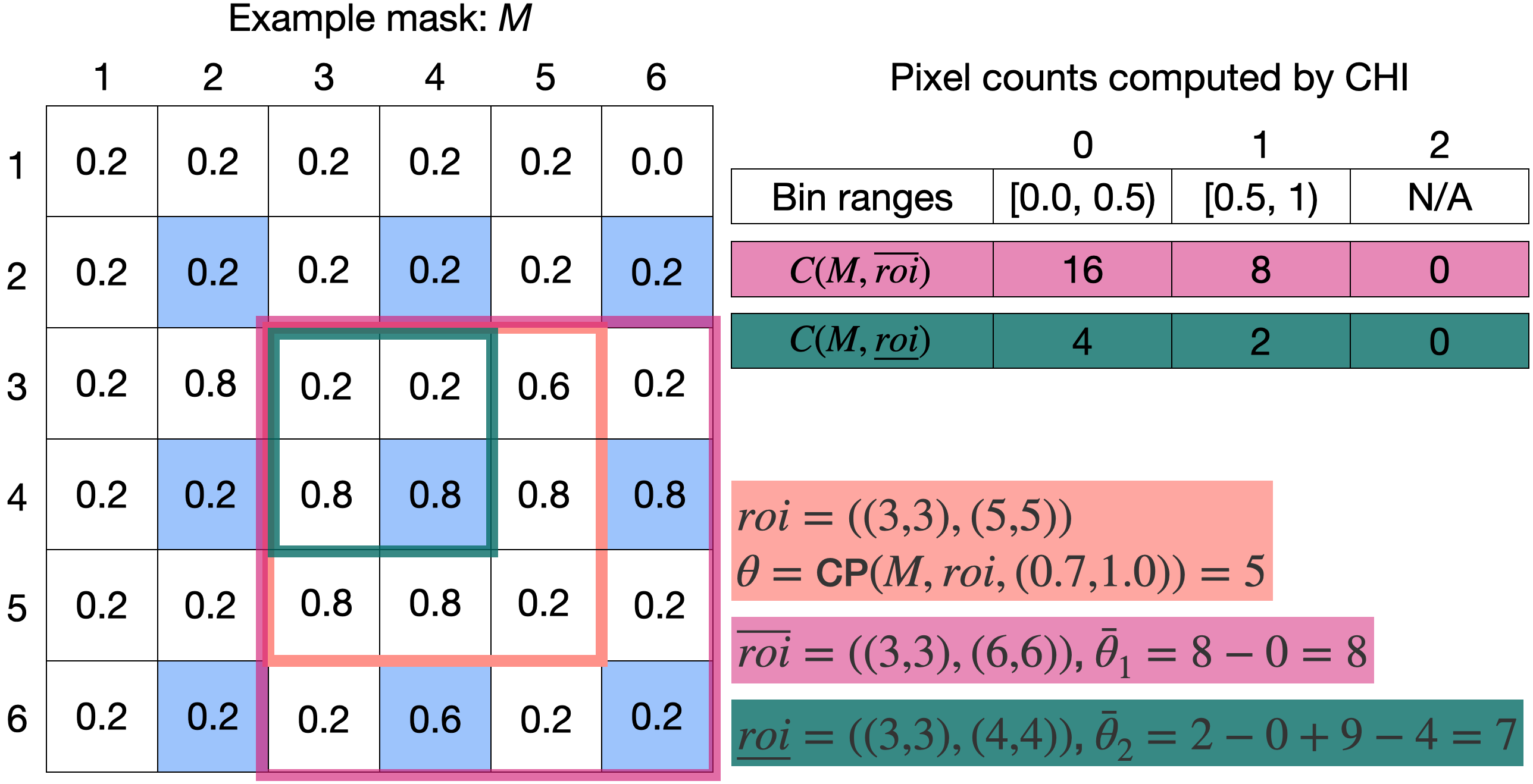}
    \vspace{-1.2em}
    \caption{An example of \system computing the upper bounds, $\bar{\theta}_1$ and $\bar{\theta}_2$, given a mask, $roi$, and $(lv, uv)$.}
    \vspace{-1.0em}
    \label{fig:upper-bound-illustration}
    \end{figure}
}

\newcommand{\additiveFunctionIllustrationFigure}{
    \begin{figure}
    \centering
    \includegraphics[width=0.85\linewidth]{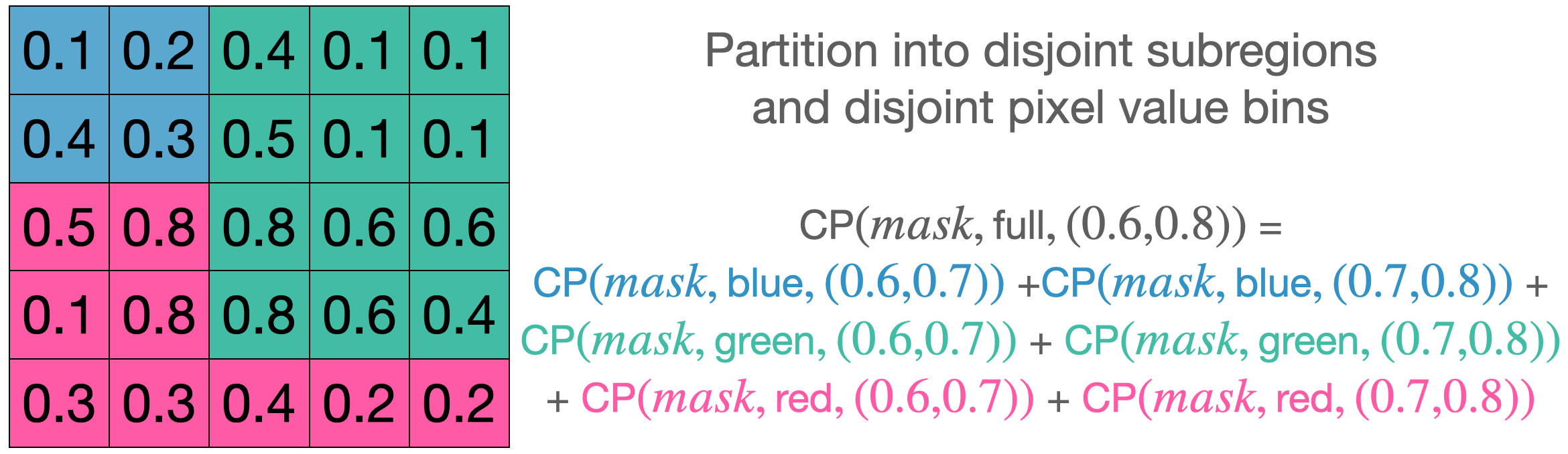}
    \vspace{-1.0em}
    \caption{Illustration of $\texttt{CP}$ being a (finitely)-additive function.}
    \vspace{-1.0em}
    \label{fig:additive-function-illustration}
    \end{figure}
}

\newcommand{\queryBasedOnMotivationTable}{
    \begingroup
    \setlength{\tabcolsep}{2pt}
    \begin{table*}[t!]
        \footnotesize
        \centering
        \caption{Summary of evaluated queries based on motivation and related work.}
        \vspace{-1.1em}
        \begin{tabularx}{\textwidth}{ l X }
            \toprule
            Query & Description \\
            \midrule
            Q1 & Returns masks s.t. $\texttt{CP}(mask, roi, (lv, uv)) > \num{5000}$, $roi = ((50, 50), (200, 200))$, $(lv, uv) = (0.6, 1.0)$, $model\_id = 1$ \\
            Q2 & Returns masks s.t. $\texttt{CP}(mask, roi, (lv, uv)) > \num{15000}$, $roi = \text{object}$, $(lv, uv) = (0.8, 1.0)$, $model\_id = 1$\\
            Q3 & Returns top-25 masks with largest $\texttt{CP}(mask, roi, (lv, uv))$, $roi = ((50, 50), (200, 200))$, $(lv, uv) = (0.8, 1.0)$, $model\_id = 1$ \\
            Q4 & Returns top-25 images with largest $\texttt{mean}(\texttt{CP}(mask, roi, (lv, uv)))$ (groupby $image\_id$) for $mask$s associated with two models, $roi = \text{object}$, $(lv, uv) = (0.8, 1.0)$ \\
            Q5 & Returns top-25 images with largest $\texttt{CP}(\texttt{intersect}(mask), roi, (lv, uv))$ (groupby $image\_id$) for $mask$s associated with two models, $roi = \text{object}$, $(lv, uv) = (0.8, 1.0)$ \\
            \bottomrule
        \end{tabularx}
        \normalsize
        \vspace{-1.0em}
        \label{tab:query-based-on-motivation}
    \end{table*}
    \endgroup
}

\newcommand{\masksLoadedTable}{
    \begingroup
    \setlength{\tabcolsep}{2pt}
    \begin{table}[t!]
        \footnotesize
        \centering
        \caption{Number of masks loaded during query execution. PG = PostgreSQL, TDB = TileDB, NP = NumPy.}
        \vspace{-1.1em}
        \begin{tabular}{ c c c c c c c c }
            \toprule
            Dataset & Method & Q1 & Q2 & Q3 & Q4 & Q5 \\
            \midrule
            \multirow{1}[2]{*}{\textit{WILDS}} & \system & 407 & 40 & 32 & 874 & 48 \\
             & PG \& TDB \& NP & \num{22275} & \num{22275} & \num{22275} & \num{44550} & \num{22275} \\
            \midrule
            \multirow{1}[2]{*}{\textit{ImageNet}} & \system & \num{2696} & \num{3849} & \num{2943} & \num{1494} & \num{2768} \\
             & PG \& TDB \& NP & \num{1331167} & \num{1331167} & \num{1331167} & \num{2662334} & \num{1331167} \\
            \bottomrule
        \end{tabular}
        \normalsize
        \vspace{-1.5em}
        \label{tab:masks-loaded}
    \end{table}
    \endgroup
}

\newcommand{\singleQueryBasedOnMotivationFigure}{
    \begin{figure*}[t!]
        \vspace{-2.3em}
        \begin{center}
            \captionsetup{font={color=\revisioncolor}}
            \hspace{1.9em}
            \subfloat{\includegraphics[width=0.60\columnwidth]{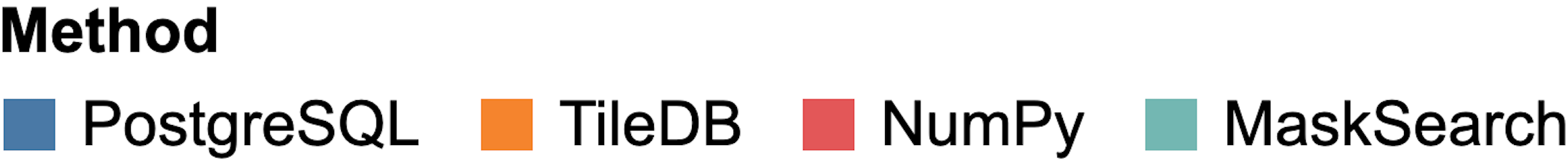}} \hfill%
            \vspace{-1.1em}
            \setcounter{subfigure}{0}

            \subfloat[\revision{\textit{WILDS}}]{\includegraphics[width=0.48\linewidth]{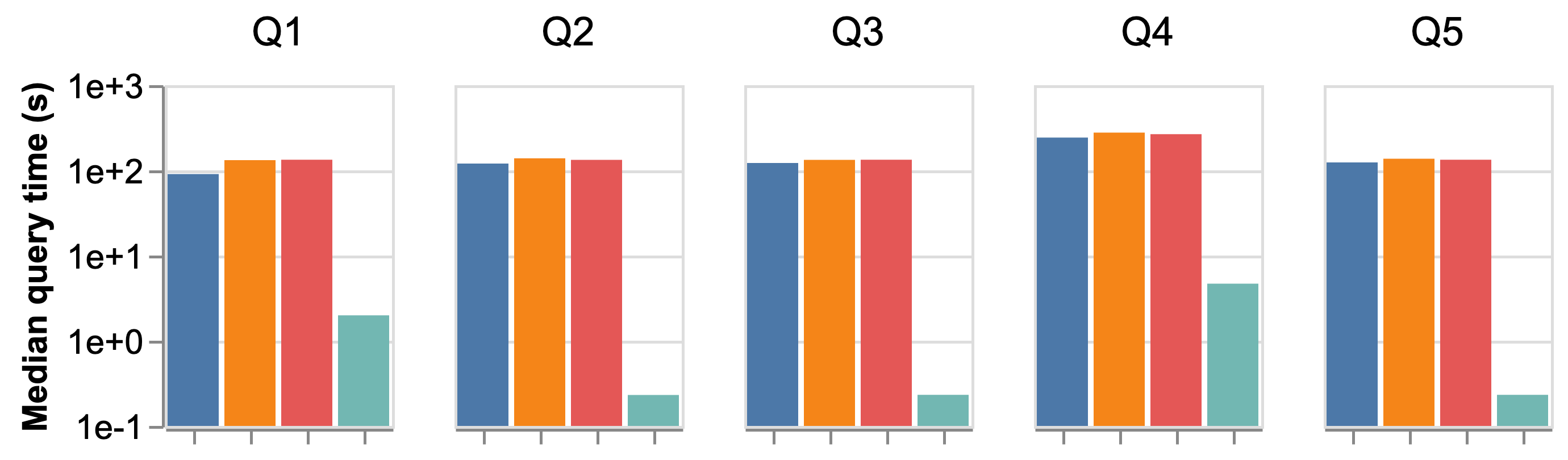}%
                \label{subfig:single-query-wilds}}%
            \hfil
            \subfloat[\revision{\textit{ImageNet}}]{\includegraphics[width=0.48\linewidth]{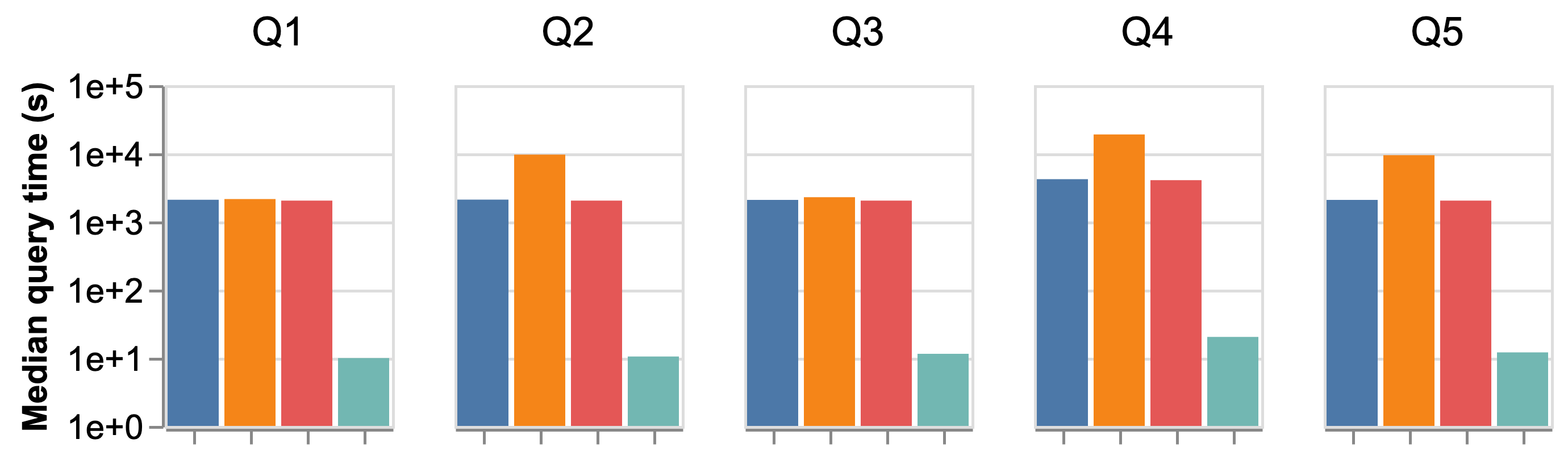}%
                \label{subfig:single-query-imagenet}}%
        \end{center}
        \vspace{-1.2em}
        \caption{End-to-end individual query execution time based on motivation and related work. The index size for \system is $\sim5\%$ of the original compressed dataset size for both datasets. Note the log scale on the y-axis.}
        \vspace{-0.8em}
        \label{fig:single-query-performance}
    \end{figure*}
}

\newcommand{\queryTimeVsQueryTypeFigure}{
    \begin{figure}[t!]
        \vspace{-1.0em}
        \begin{center}

            \subfloat[\revision{\textit{WILDS}}]{\includegraphics[width=0.5\linewidth]{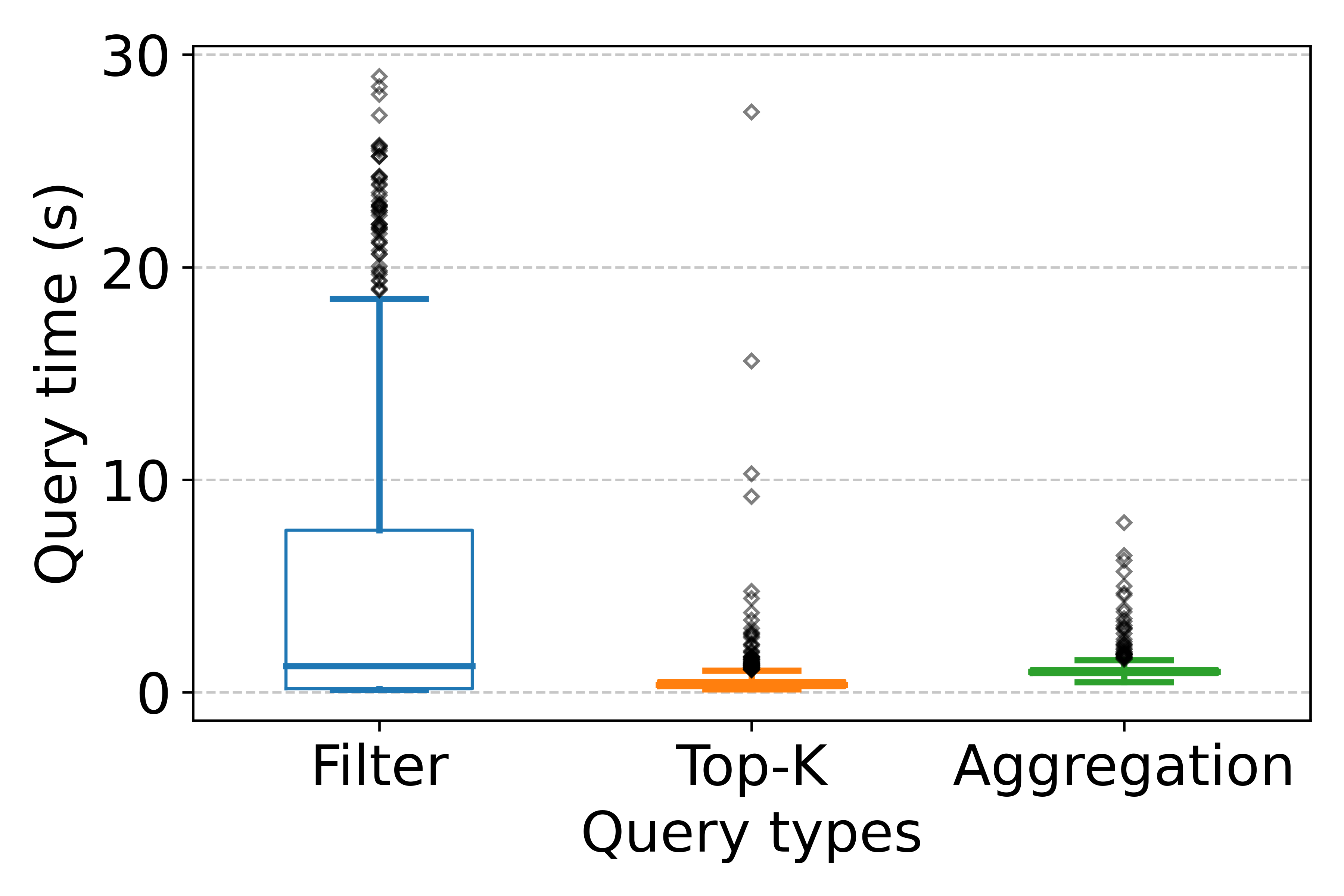}%
                \label{subfig:query-time-vs-query-type-wilds}}%
            \hfil
            \subfloat[\revision{\textit{ImageNet}}]{\includegraphics[width=0.5\linewidth]{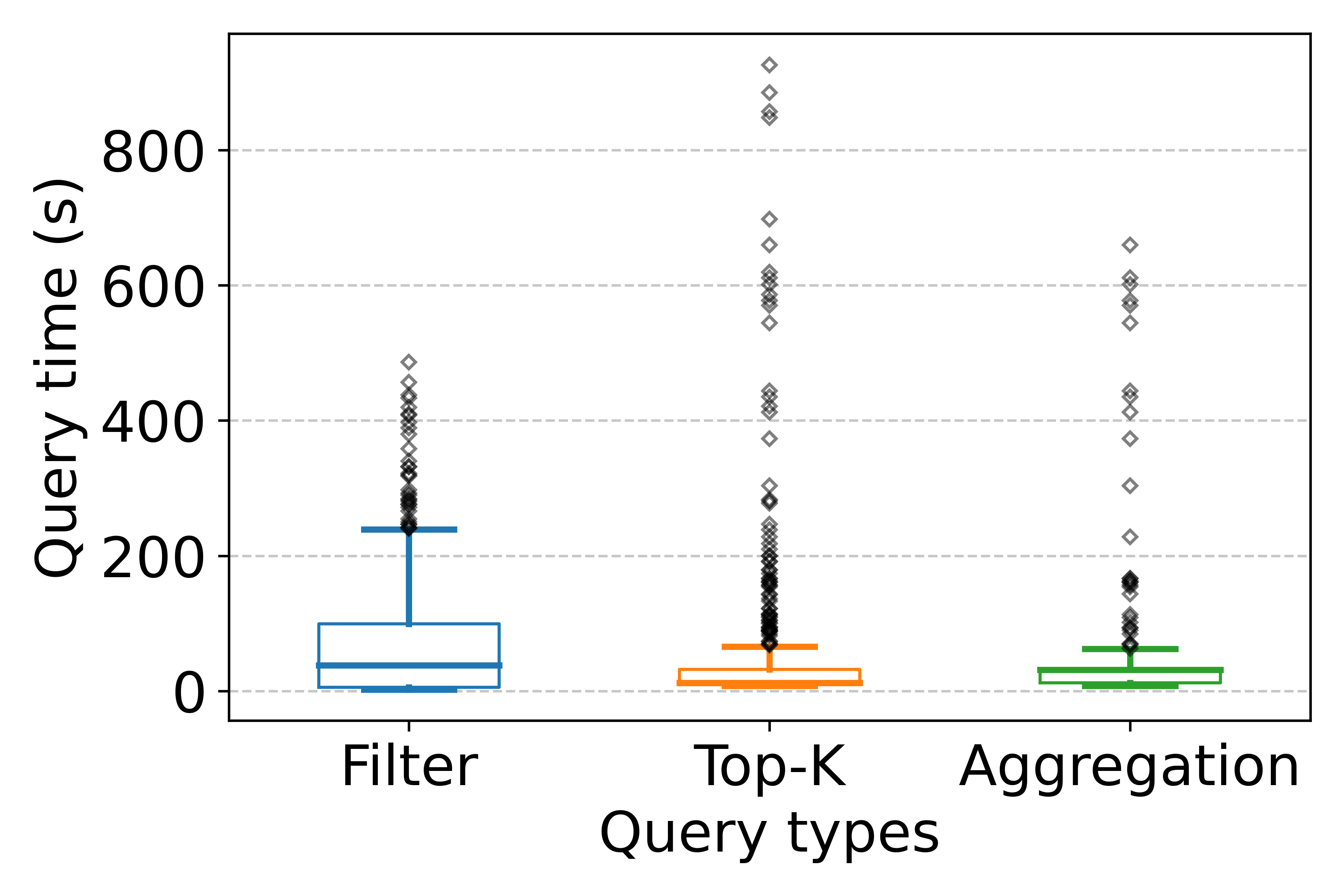}%
                \label{subfig:query-time-vs-query-type-imagenet}}%
        \end{center}
        \vspace{-1.2em}
        \caption{Query time of \system for different query types. Index size for \system: $\sim5\%$ of dataset size.}
        \vspace{-1em}
        \label{fig:query-time-vs-query-type}
    \end{figure}
}

\newcommand{\queryTimeVsFractionOfMasksLoadedFigure}{
    \begin{figure}[t!]
        \begin{center}

            \subfloat[\textit{WILDS}, Pearson's $r = 0.99$]{\includegraphics[width=0.48\linewidth]{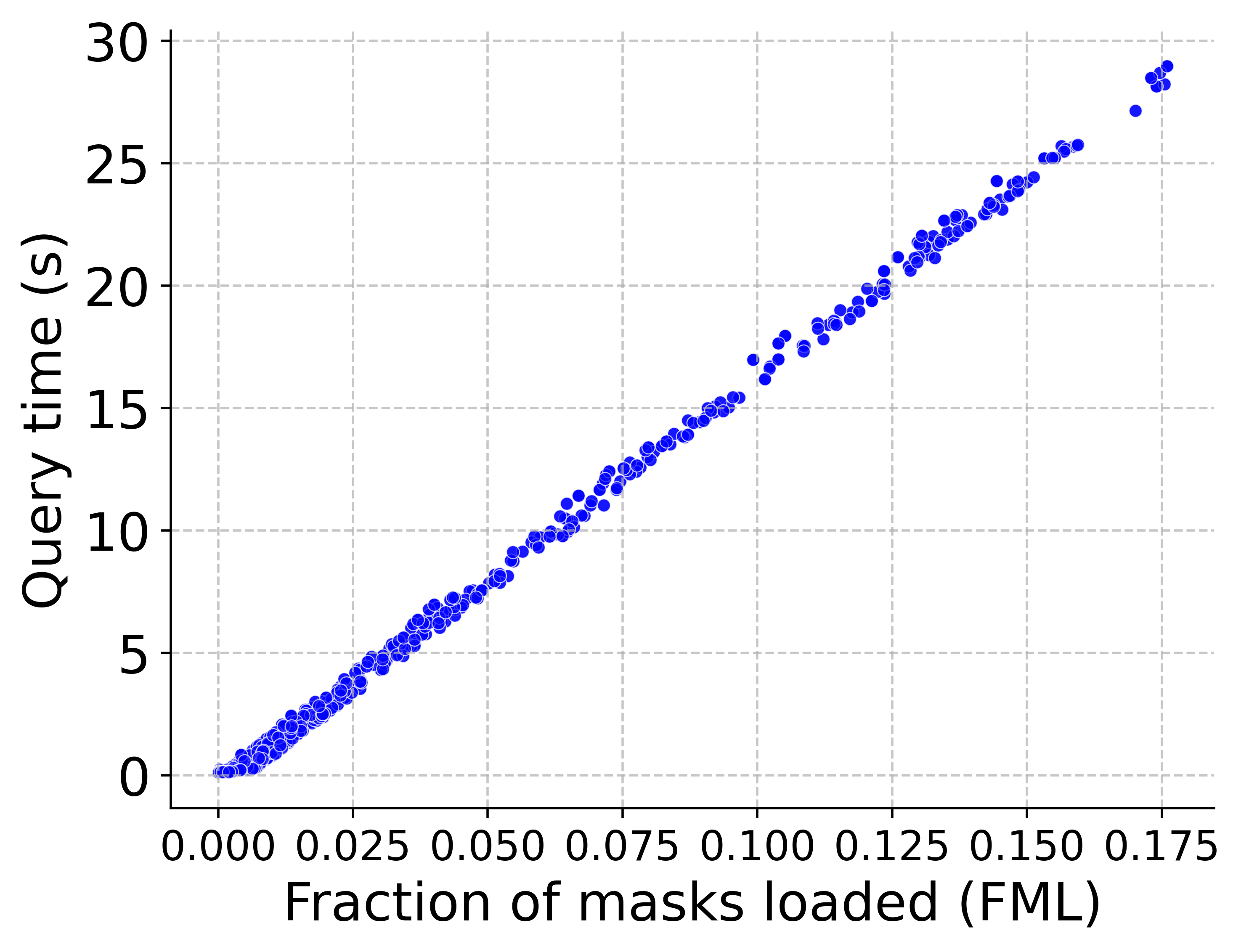}%
                \label{subfig:query-time-vs-fraction-of-masks-loaded-wilds}}%
            \hfil
            \subfloat[\textit{ImageNet}, Pearson's $r = 0.96$]{\includegraphics[width=0.5\linewidth]{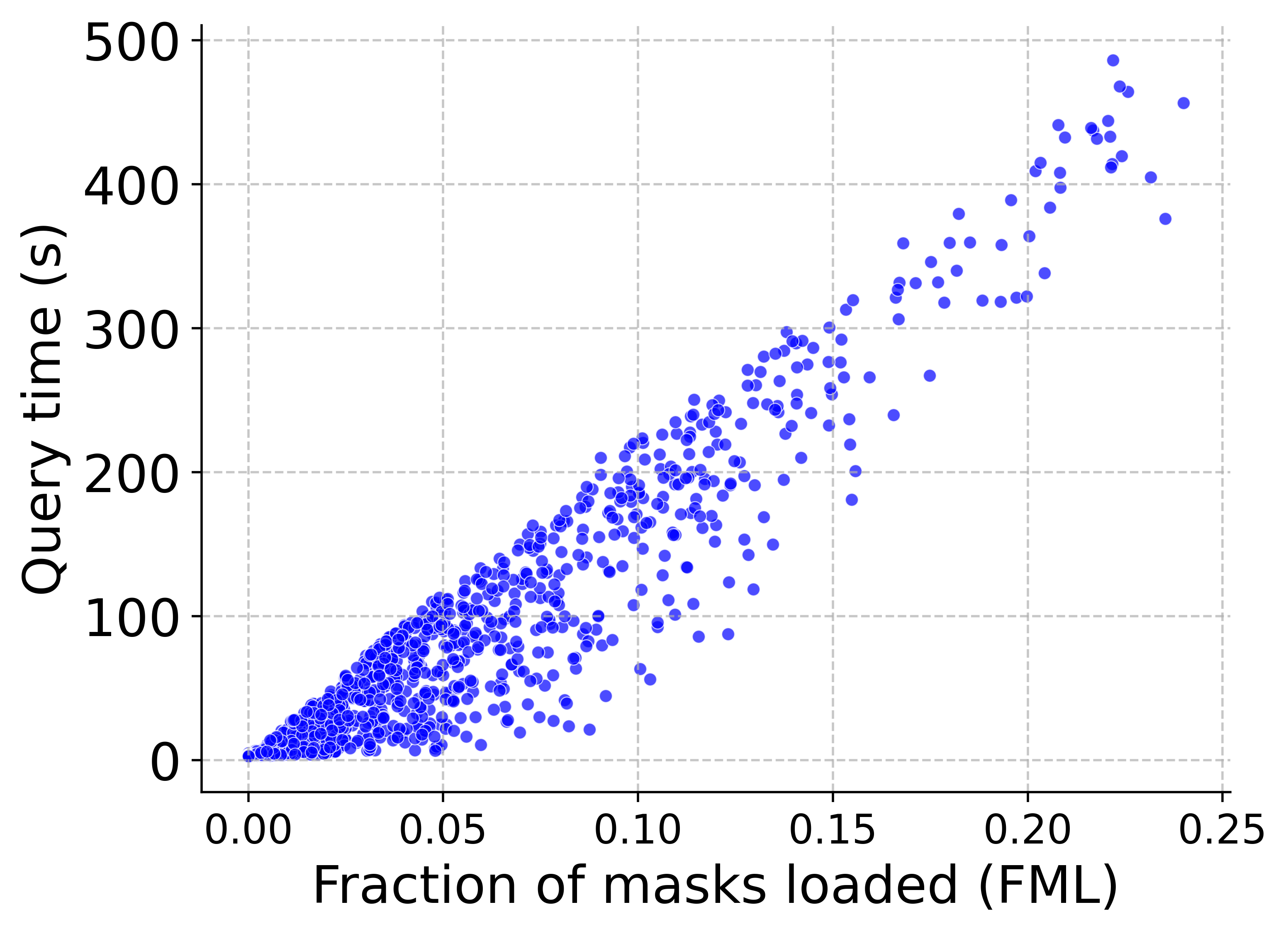}%
                \label{subfig:query-time-vs-fraction-of-masks-loaded-imagenet}}%
        \end{center}
        \vspace{-1.2em}
        \caption{Relationship between end-to-end query time and the fraction of masks loaded (FML) for a query.}
        \vspace{-1em}
        \label{fig:query-time-vs-fraction-of-masks-loaded}
    \end{figure}
}

\newcommand{\combinedBoundSegmentsFigure}{
    \begin{figure*}[t!]
        \begin{center}
            \subfloat[\textit{WILDS, 88 MB, $(0.6, 1.0)$}]{\includegraphics[width=0.25\linewidth]{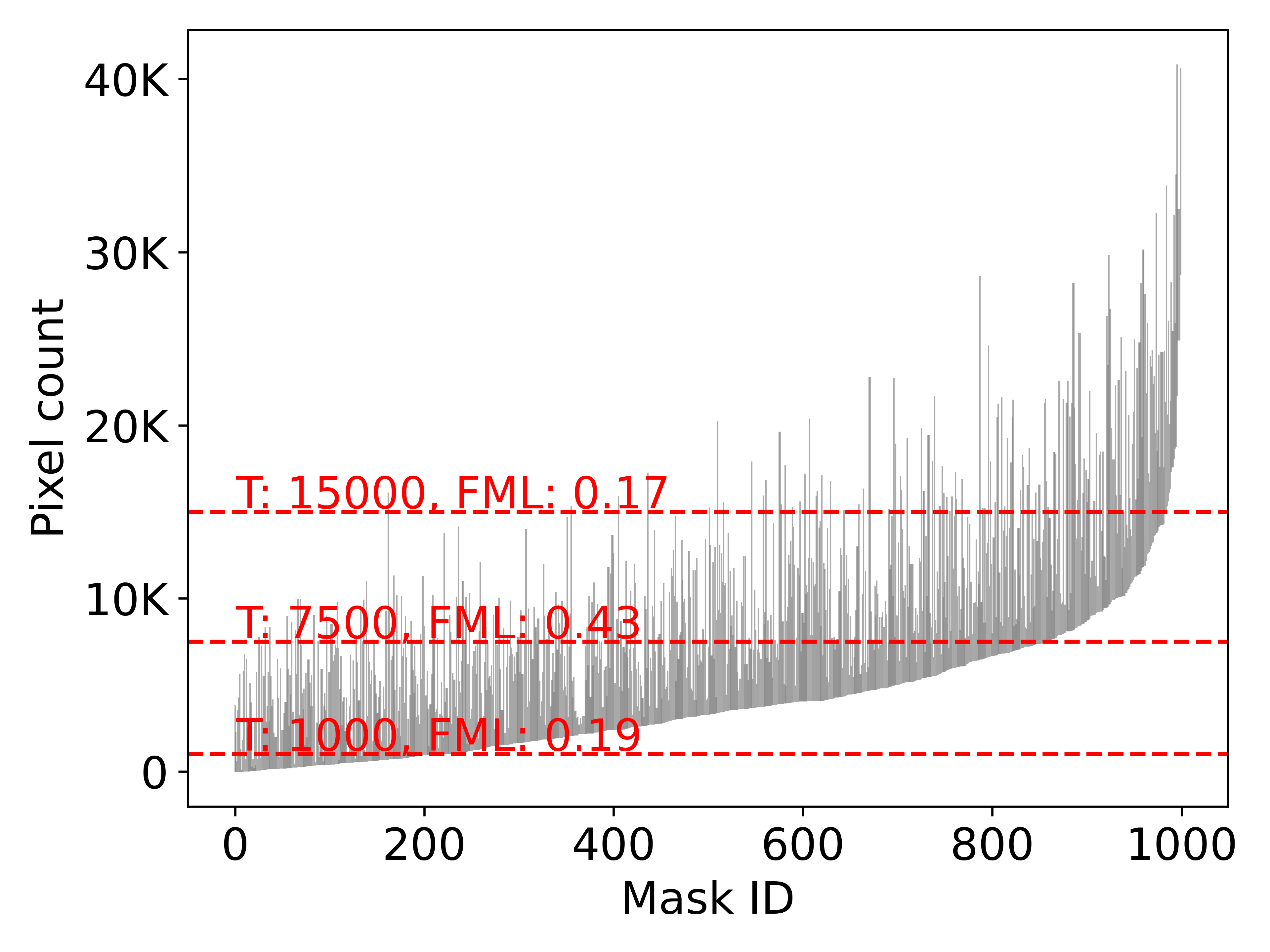}\vspace{-0.8em}}%
            \hfil
            \subfloat[\textit{WILDS, 88 MB, $(0.8, 1.0)$}]{\includegraphics[width=0.25\linewidth]{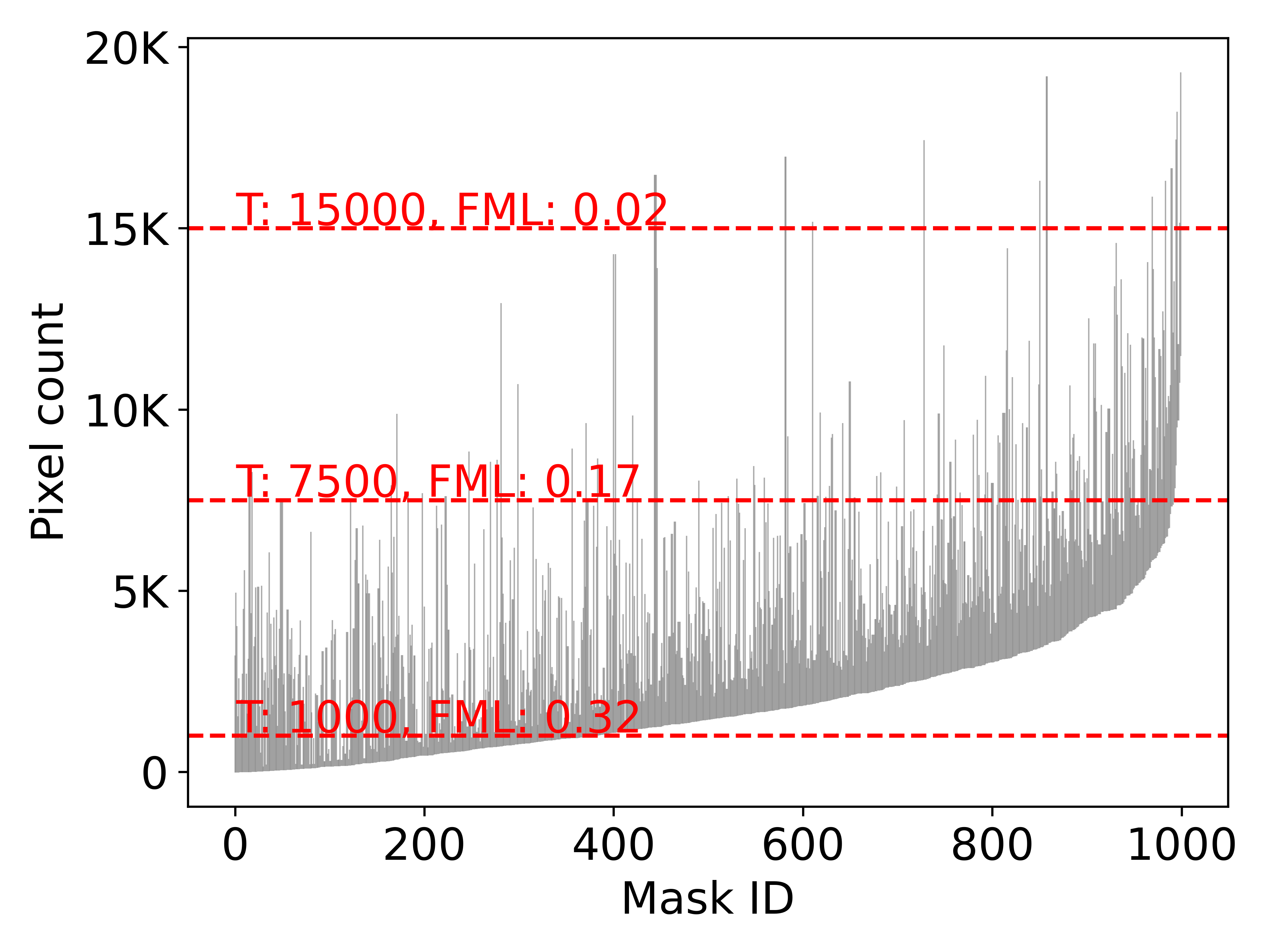}\vspace{-0.8em}}%
            \hfil
            \subfloat[\textit{WILDS, 2.2 GB, $(0.6, 1.0)$}]{\includegraphics[width=0.25\linewidth]{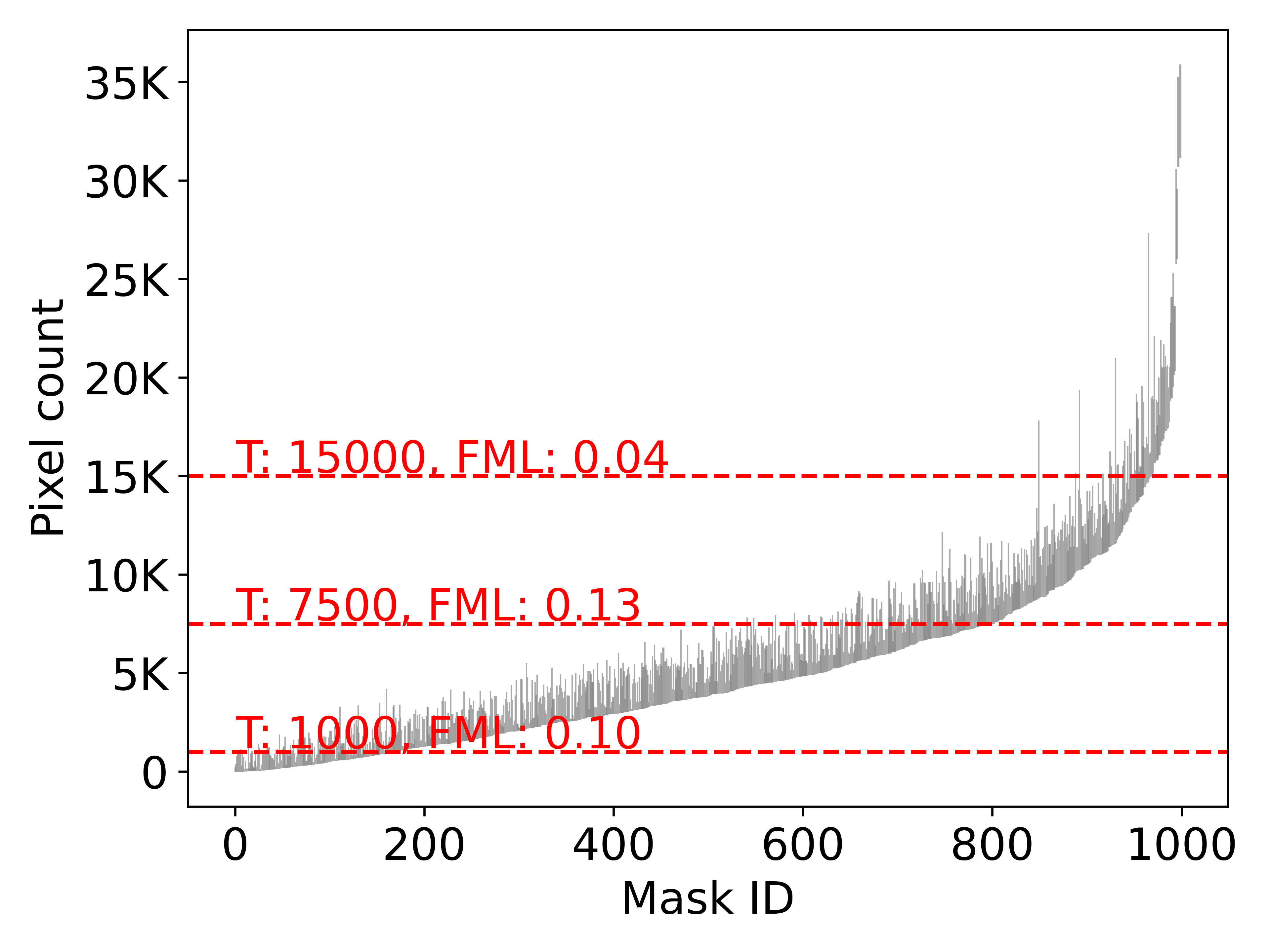}\vspace{-0.8em}}%
            \hfil
            \subfloat[\textit{WILDS, 2.2 GB, $(0.8, 1.0)$}]{\includegraphics[width=0.25\linewidth]{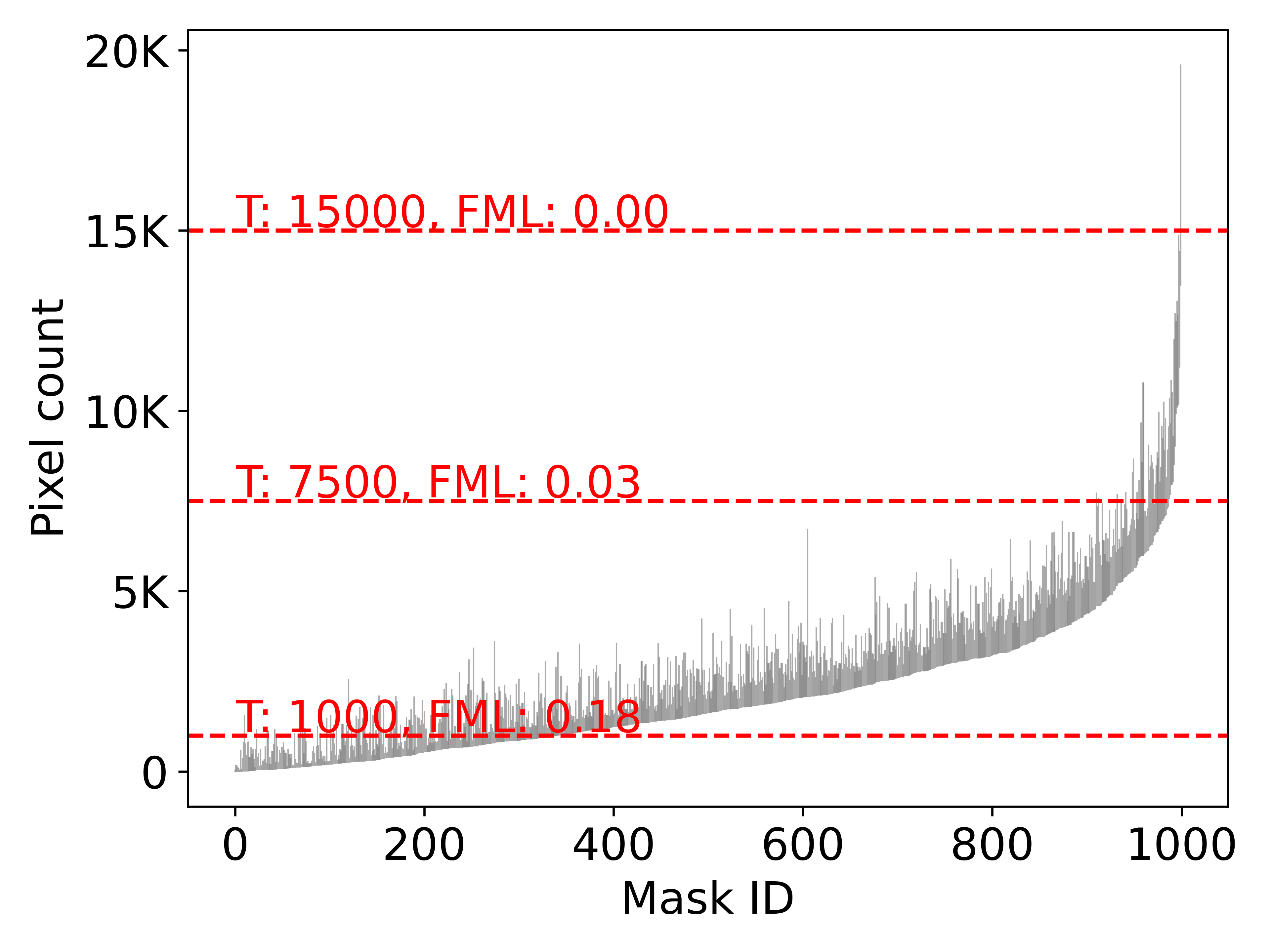}\vspace{-0.8em}}%
            \vspace{-1.1em}
            \subfloat[\textit{ImageNet, 6.5 GB, $(0.6, 1.0)$}]{\includegraphics[width=0.25\linewidth]{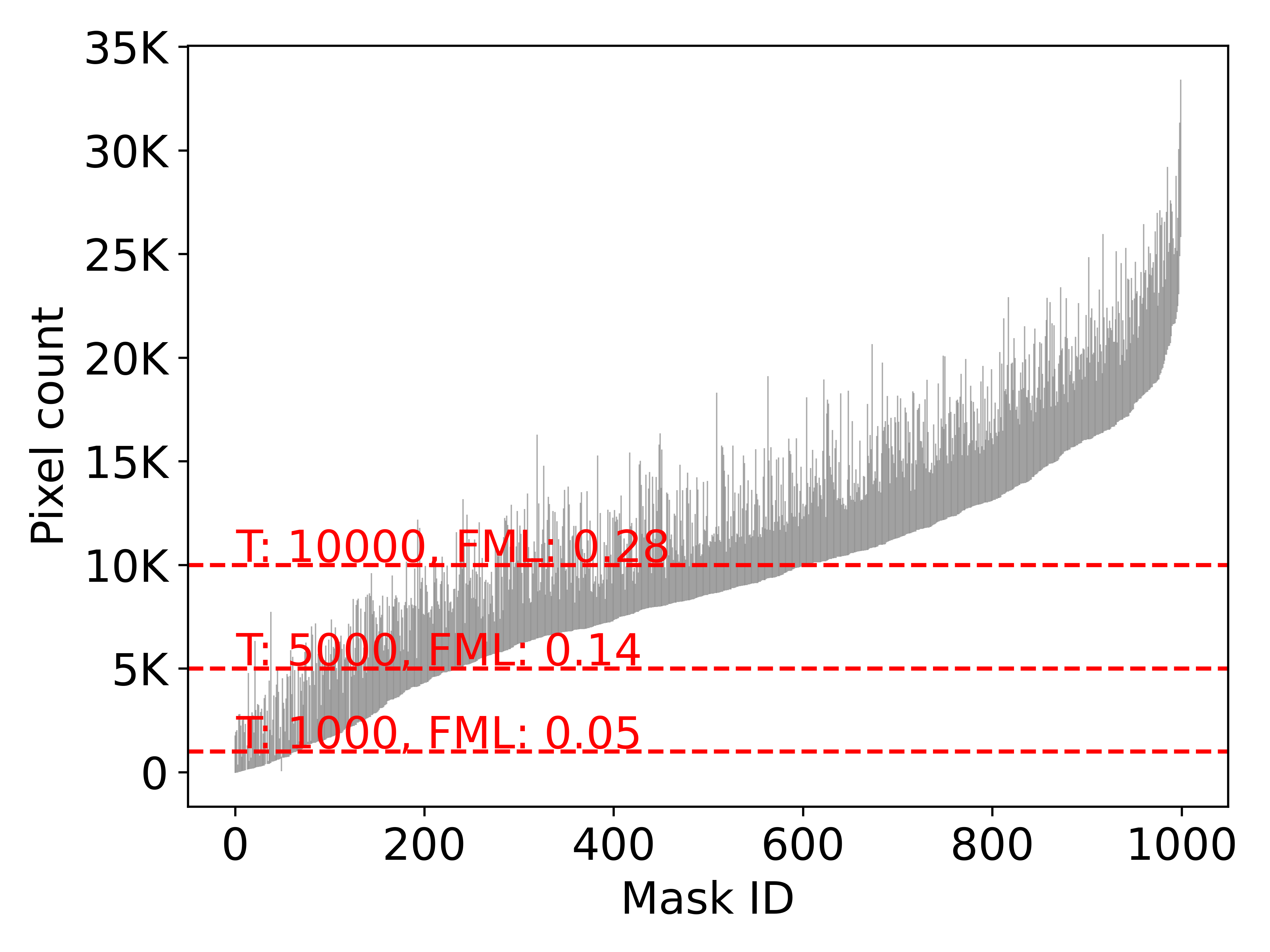}\vspace{-0.8em}}%
            \hfil
            \subfloat[\textit{ImageNet, 6.5 GB, $(0.8, 1.0)$}]{\includegraphics[width=0.25\linewidth]{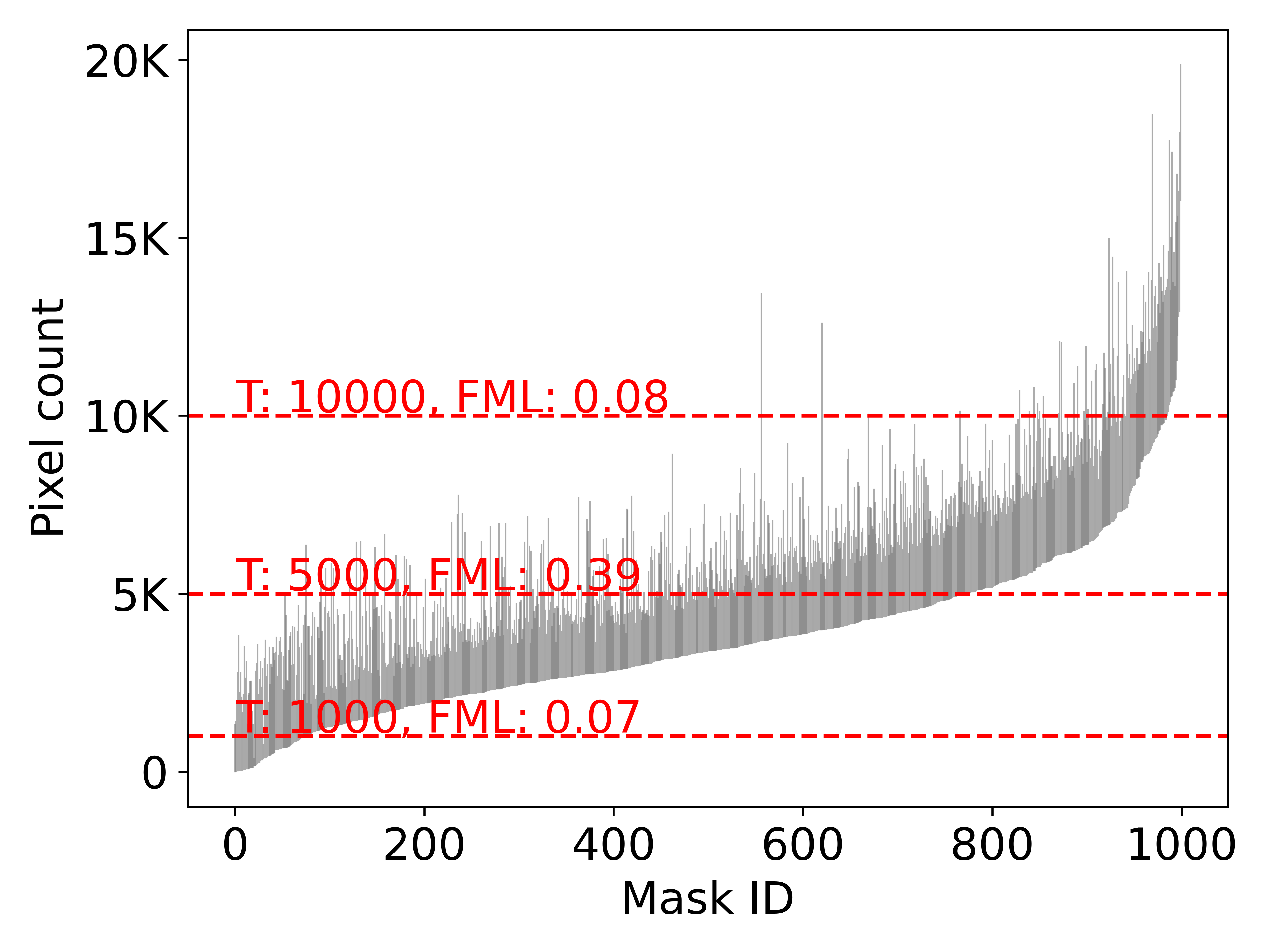}\vspace{-0.8em}}%
            \hfil
            \subfloat[\textit{ImageNet, 23 GB, $(0.6, 1.0)$}]{\includegraphics[width=0.25\linewidth]{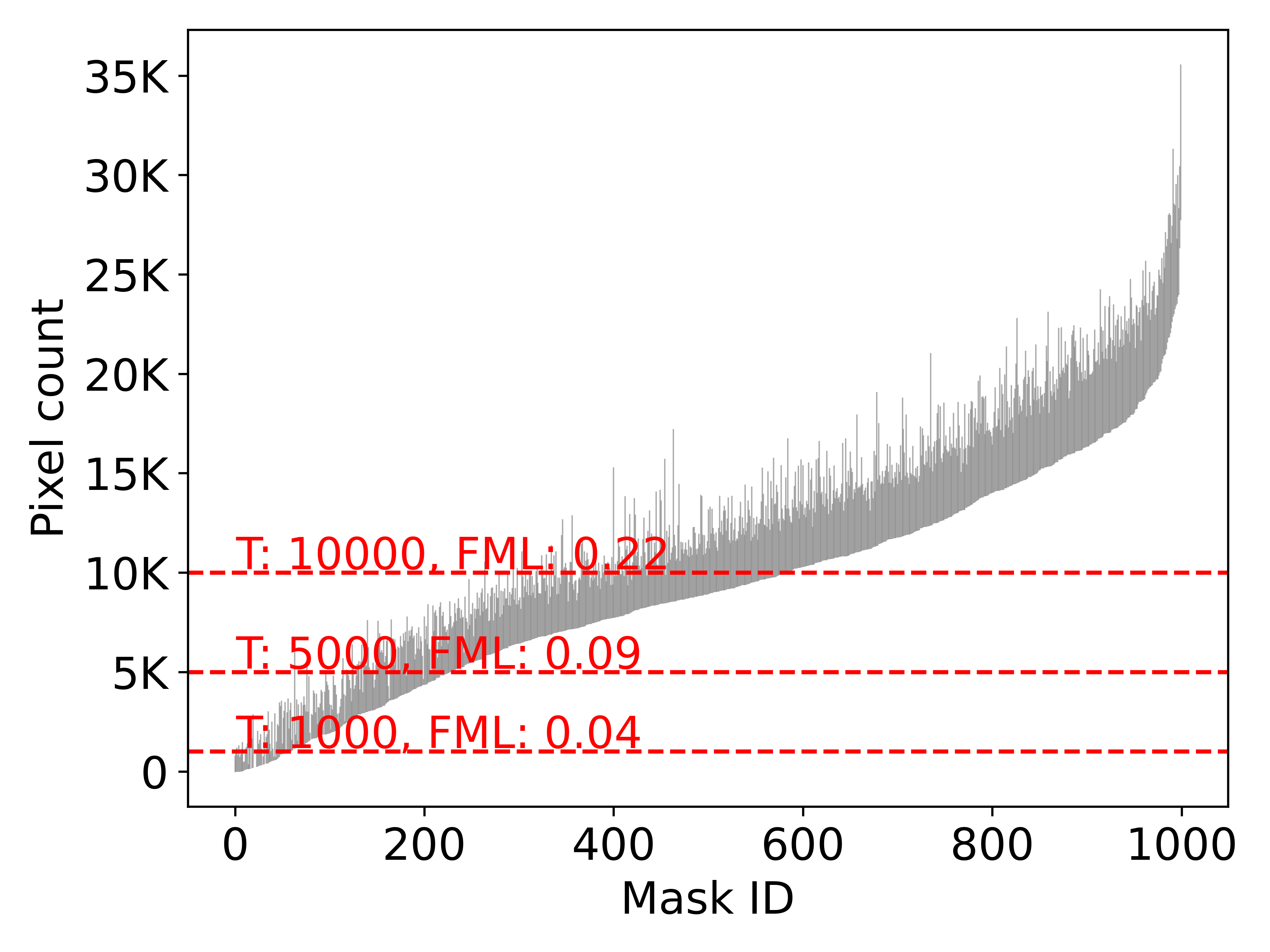}\vspace{-0.8em}}%
            \hfil
            \subfloat[\textit{ImageNet, 23 GB, $(0.8, 1.0)$}]{\includegraphics[width=0.25\linewidth]{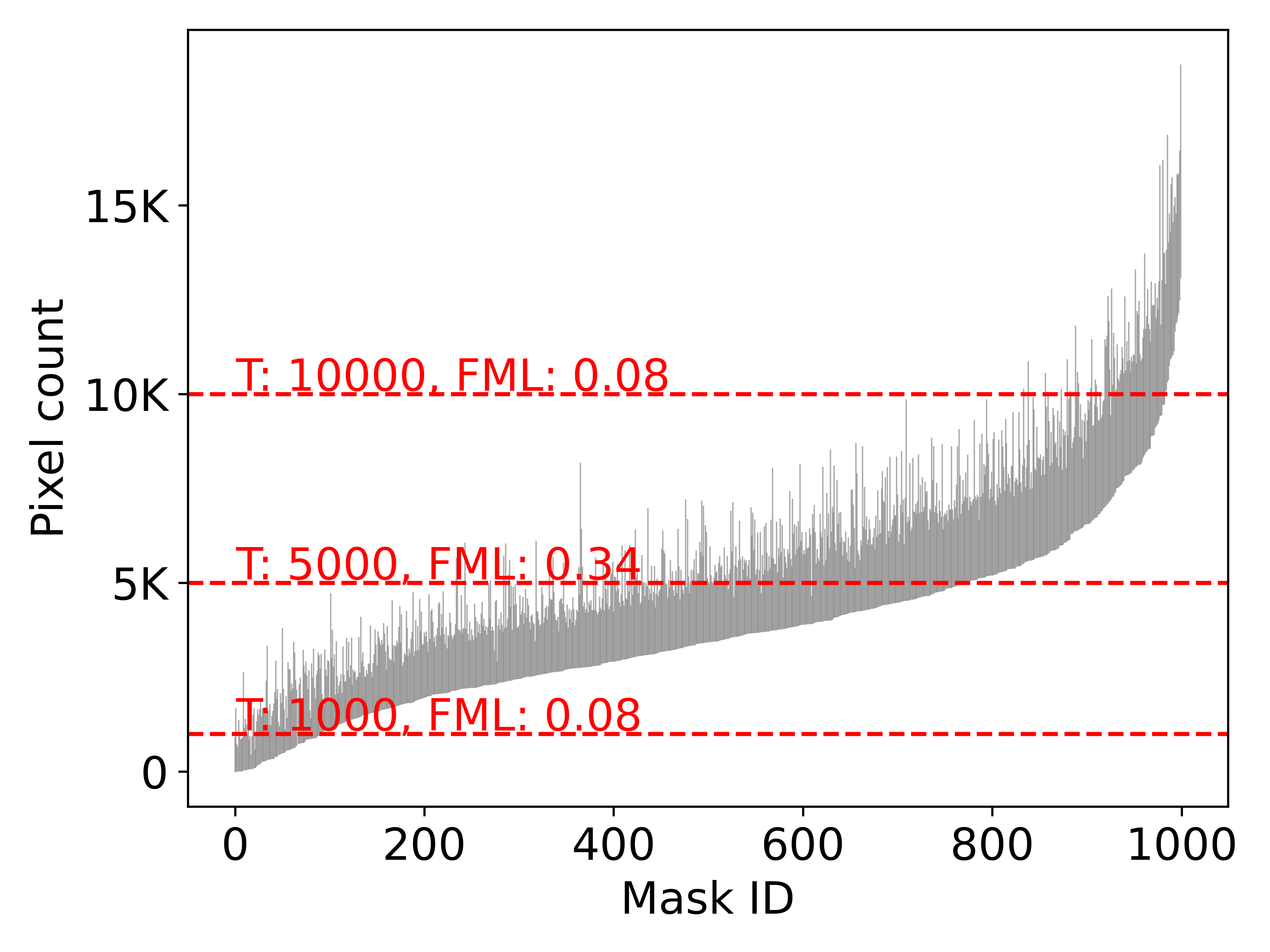}\vspace{-0.8em}}%
        \end{center}
        \vspace{-1.2em}

        \caption{Distribution of bounds of $\texttt{CP}(mask, roi, (lv, uv))$ computed by \system. Each subfigure represents the distribution for a combination of $(\text{dataset}, \text{index size}, (lv, uv))$, shown as the title of each. Each vertical segment represents the lower and upper bounds of $\texttt{CP}(mask, roi, (lv, uv))$ for a single mask. For each mask, $roi$ is the foreground object bounding box. We show the distribution of bounds for \num{1000} randomly sampled masks in each subplot. The x-axes represent the masks sorted by their lower bounds. The horizontal dashed lines represent examples of the count threshold $T$. FML is the fraction of masks loaded by \system given a predicate $\texttt{CP}(mask, roi, (lv, uv)) > T$. For each count threshold $T$, FML is equal to the fraction of the vertical segments that intersect with the horizontal dashed line defined by $T$. Note the different scales of the y-axes.}
        \vspace{-1.0em}
        \label{fig:combined-bound-segments}
    \end{figure*}
}

\newcommand{\multiQueryWorkloadFigure}{
    \begin{figure*}[t!]
        \begin{center}
            \captionsetup{font={color=black}}
            \hspace{1em}
            \subfloat{\includegraphics[width=0.6225\columnwidth]{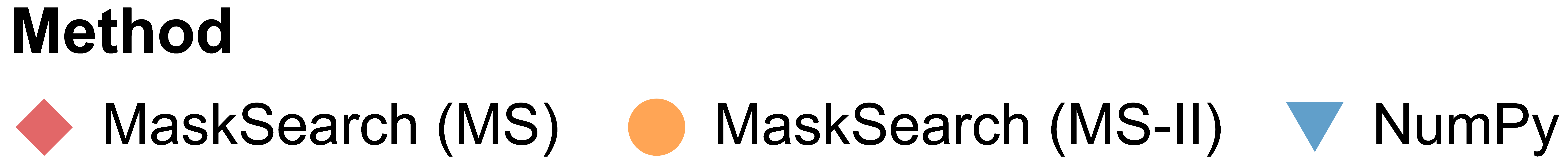}}%
            \setcounter{subfigure}{0}
            \hfil
            \hspace{11.50em}
            \subfloat{\includegraphics[width=0.660\columnwidth]{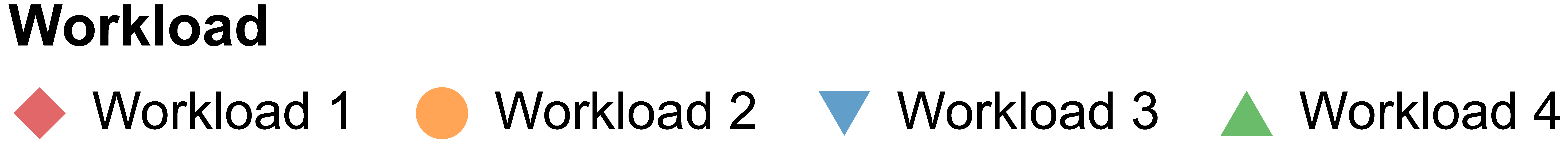}} \hfill%
            \vspace{-1.1em}
            \setcounter{subfigure}{0}

            \subfloat[\revision{\textit{WILDS, Workload 2}}]{\includegraphics[width=0.24\linewidth]{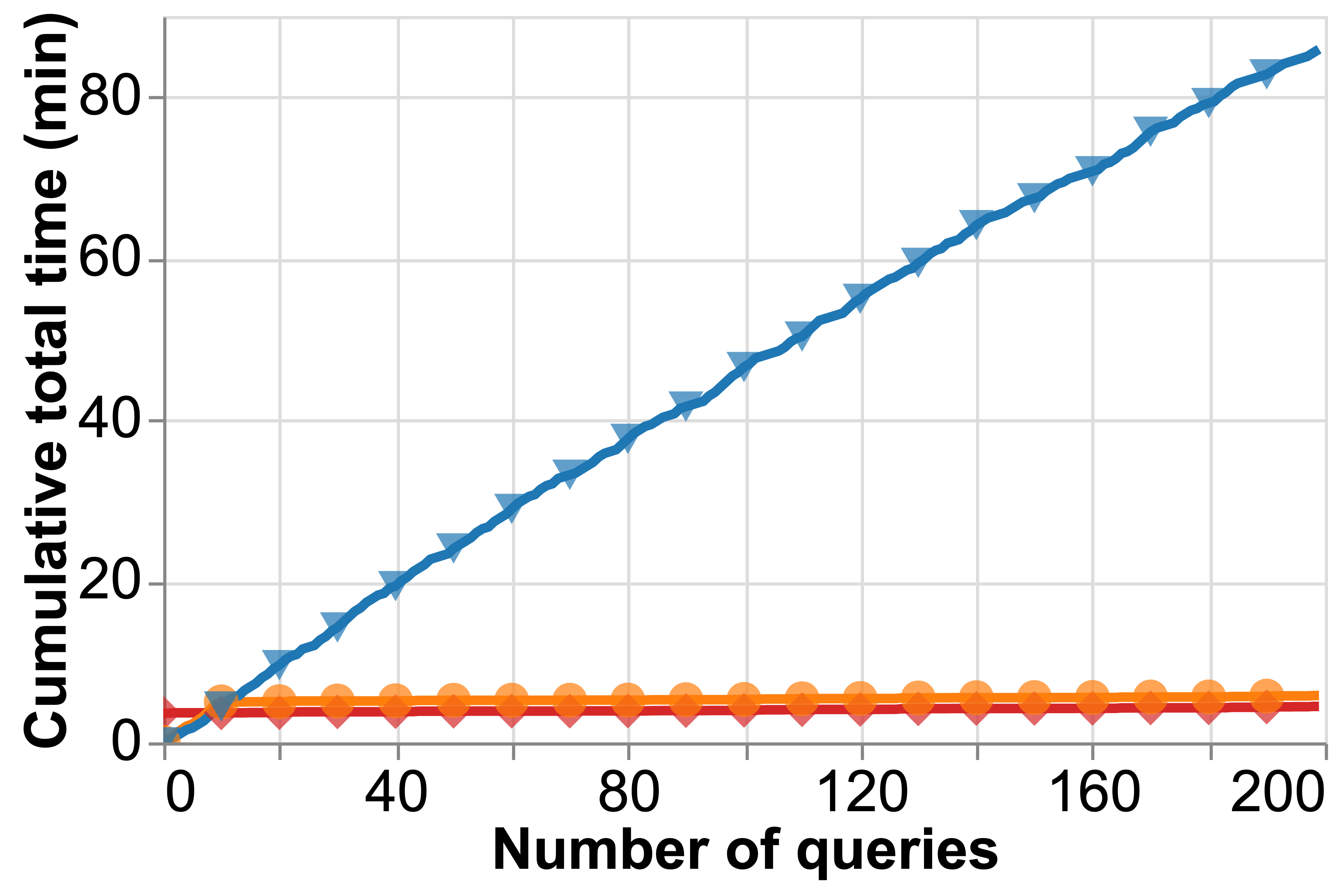}%
                \label{subfig:multi-query-workload-a}}%
            \hfil
            \subfloat[\revision{\textit{ImageNet, Workload 2}}]{\includegraphics[width=0.24\linewidth]{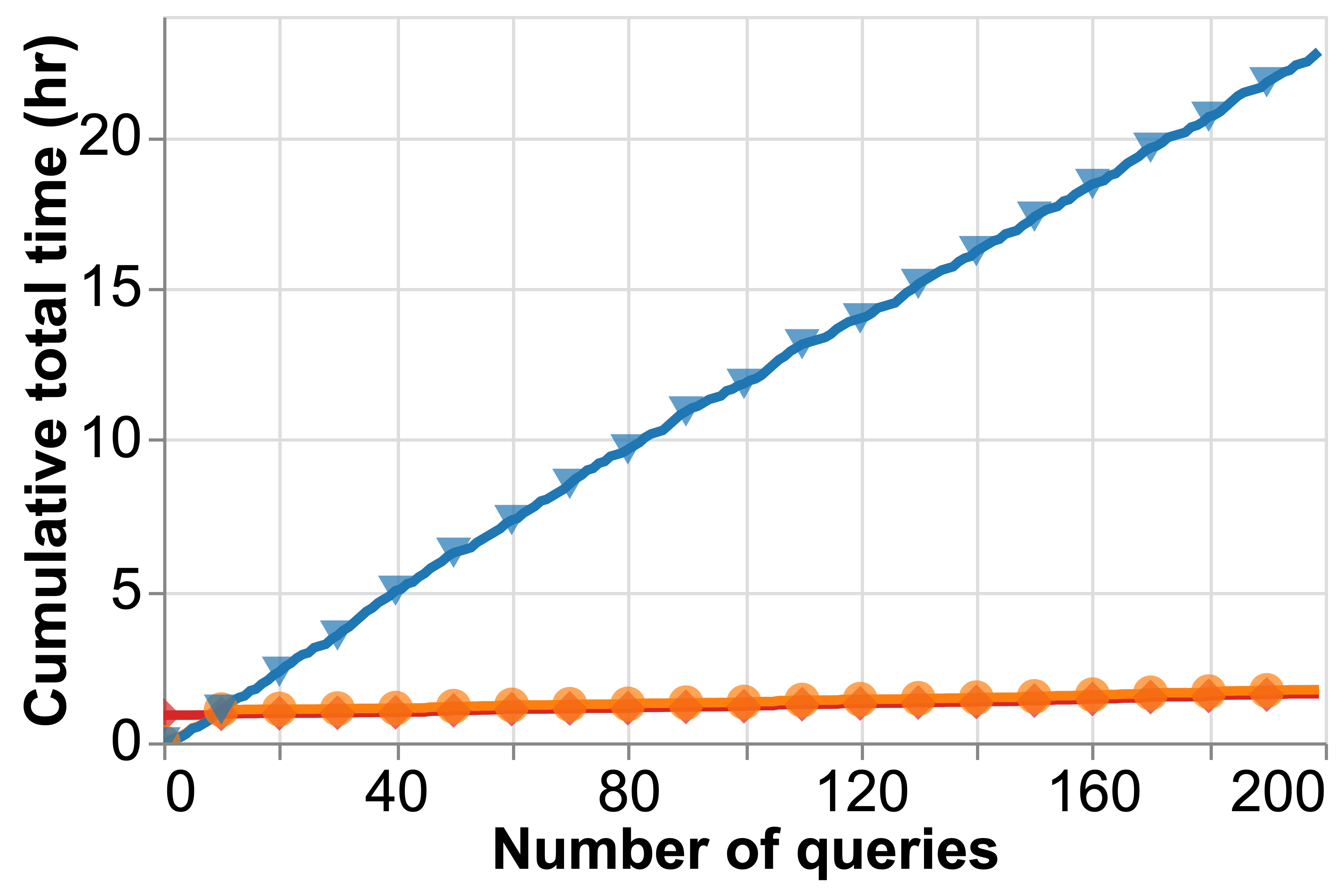}%
                \label{subfig:multi-query-workload-b}}%
            \hfil
            \subfloat[\revision{\textit{WILDS, MS-II vs. MS}}]{\includegraphics[width=0.24\linewidth]{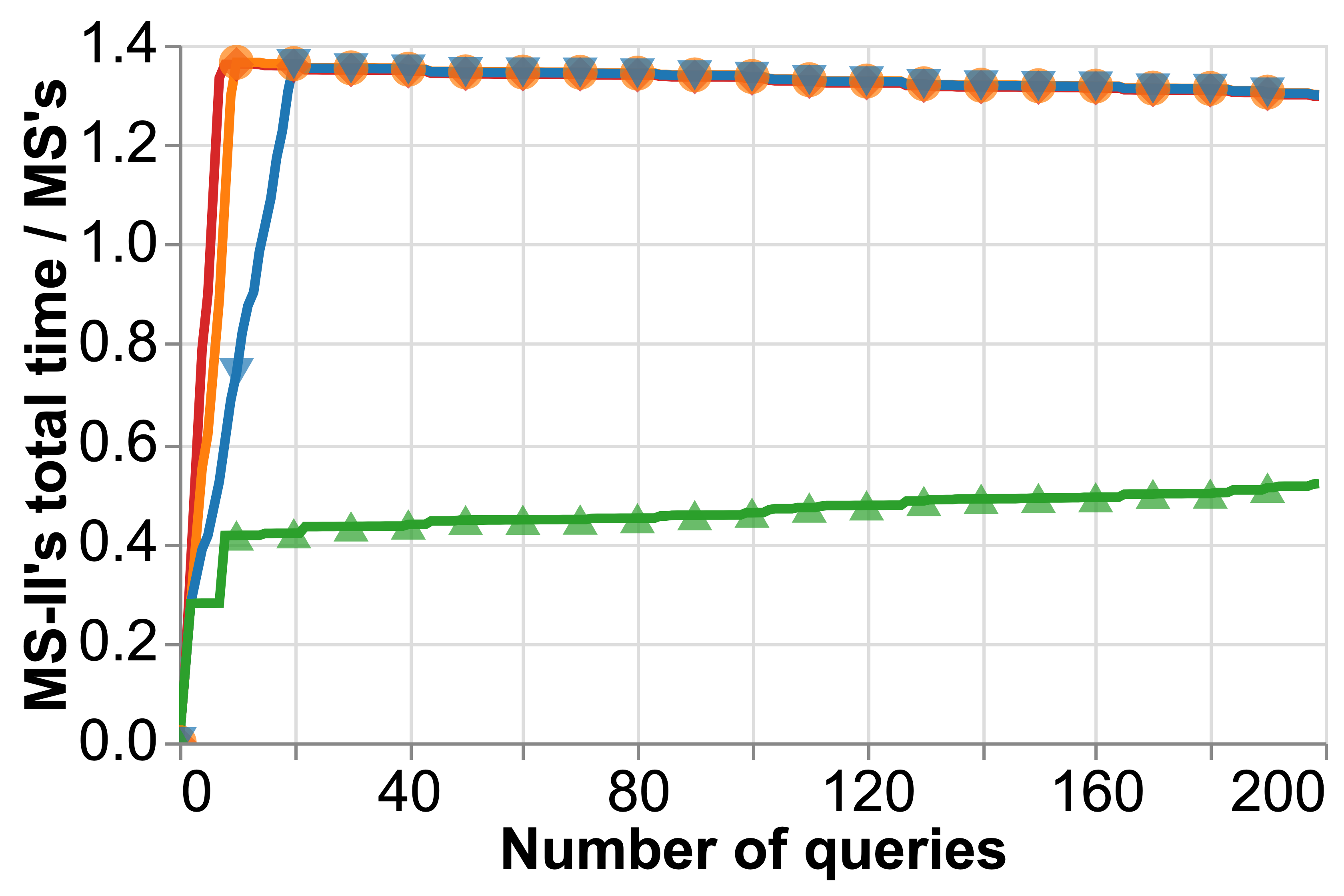}%
                \label{subfig:multi-query-workload-c}}%
            \hfil
            \subfloat[\revision{\textit{ImageNet, MS-II vs. MS}}]{\includegraphics[width=0.24\linewidth]{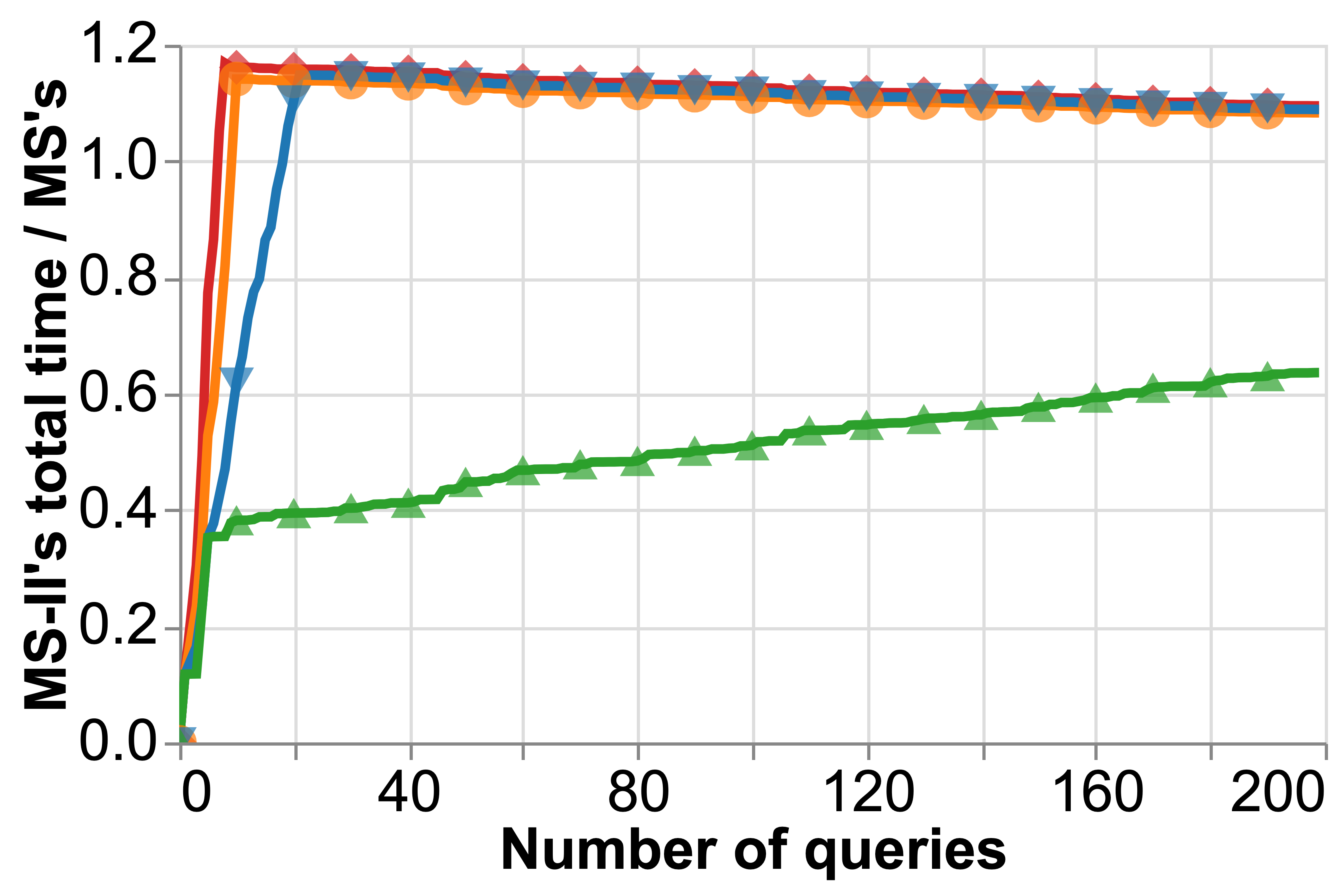}%
                \label{subfig:multi-query-workload-c}}%
        \end{center}
        \vspace{-1.0em}
        \caption{Cumulative total time, incl. index building time and query time, for multi-query workloads. MS-II and MS refer to \system w/ and w/o incremental indexing, respectively. (a) and (b) show the total time for MS, MS-II, and NumPy for Workload 2; (c) and (d) show the ratio of the cumulative total time of MS-II to that of MS for all workloads. The index size for MS is $\sim5\%$ of the corresponding dataset. MS-II builds the index incrementally using the same index configuration as MS.}
        \label{fig:multi-query-workload}
    \end{figure*}
}

\section{Introduction}
\label{sec:introduction}

Many machine learning (ML) tasks over image databases commonly generate masks that annotate individual pixels in images. 
For instance, model explanation techniques~\cite{sundararajan2017axiomatic, smilkov2017smoothgrad, gradcam2017, zhou2015cnnlocalization, singla2019understanding} generate saliency maps to highlight the significance of individual pixels to a model's output. 
In image segmentation tasks~\cite{he2018mask, kirillov2023segment, ronneberger2015unet}, masks denote the probability of pixels being associated with a specific class or an instance. 
Depth estimation models~\cite{Bhattacharjee_2022_WACV, P3Depth} yield masks reflecting the depth of each pixel, while human pose estimation models~\cite{openpose, guler2018densepose} provide masks indicating the probability of pixels corresponding to body joints. \cref{fig:mask-example} shows some examples.

Exploring the properties of these masks unlocks a plethora of applications. 
For instance, in the context of model explanation, examining saliency maps is the most common approach to understanding whether a model is relying on spurious correlations in the input data, i.e., signals that deviate from domain knowledge~\cite{oakden2020hidden, plumb2022finding, bissoto2019constructing, winkler2019association, degrave2021ai, ming2021impact}. 
Other applications based on the properties of masks include identifying maliciously attacked examples using saliency maps~\cite{ye2020detection, wang2022adversarial, zhang2018detecting}, out-of-distribution detection also using saliency maps~\cite{hornauer2022heatmapbased}, monitoring model errors~\cite{meerkat2023goel, kangdata, tesla2020gritti} using segmentation masks, traffic monitoring and retail analytics using segmentation masks~\cite{datafromsky-traffic, datafromsky-retail}, and others.

The wide-ranging applications underscore an emerging necessity for ML practitioners: the capability to efficiently query and retrieve examples from image databases together with their masks, based on properties of the latter~\cite{plumb2022finding, das2016human, kirillov2023segment}. Today, ML practitioners lack a system that would support this task efficiently and at scale.

Consider the following two scenarios inspired by the literature:

\maskExampleFigure
\introExampleFigure

\textit{\textbf{Scenario 1} (inspired by~\cite{ye2020detection}): Bob is a data engineer who is responsible for monitoring the performance of an image classification model. 
He notices a significant drop in the model's accuracy over the past week. 
To understand why, Bob examines the saliency maps for the misclassified images and finds that the high-value pixels are not concentrated on the foreground objects, but rather diffused across irrelevant background regions. 
\cref{fig:intro-example} shows three example images overlaid with their saliency maps. 
He suspects that these misclassifications might be due to malicious modifications that mislead the model to focus on irrelevant pixels. 
Bob wishes to identify and retrieve other images where high-value pixels are dispersed across large fractions of images. 
By analyzing these examples, he could better understand the extent of the malicious modifications and work towards improving the model's resilience to such attacks. }

\textit{\textbf{Scenario 2} (inspired by~\cite{degrave2021ai}): Alice is a scientist who is developing a model to detect COVID-19 based on chest X-ray images. 
She has trained a model that achieves high accuracy on both the training and validation sets from a public dataset. 
However, when the model is deployed to local hospitals, the model's prediction often contradicts the diagnosis based on PCR tests. 
Eager to understand why her high-accuracy model is failing in real-world settings, Alice examines the saliency maps generated by the model for the chest X-ray images from the training set.
She discovers that the high-value pixels in the saliency maps are concentrated on the markers around the peripheries instead of the lung regions. 
This observation suggests that the model is learning the confounding factors in the images (i.e., the lateral markers) rather than the medical pathology of the lungs.
Figure 3 in~\cite{degrave2021ai} shows example X-rays with their saliency maps that exhibit this phenomenon. 
To further investigate, Alice wishes to retrieve more examples that exhibit similar mask properties.}

As the above examples illustrate, querying databases of masks is important in ML applications. Unfortunately, 
there is a lack of system support to efficiently execute these queries~\cite{hong2020human}. 
According to~\cite{plumb2022finding}, to identify examples for which the model relies on spurious correlations, researchers had to manually examine the explanation maps for each image. 
This tedious approach is clearly untenable and calls for a system that efficiently supports mask-based queries. 

In light of existing challenges, we propose \system, a system that efficiently retrieves examples based on mask properties. 
To build \system, we first formalize a novel, and broadly applicable, class of queries that retrieve images (and their masks) from image databases based on the properties of masks computed over those images. 
At the core of these queries are predicates on image masks that apply filters and aggregations (i.e., count of pixels) on the values of pixels within regions of interest (ROIs). 
We further extend the queries to support aggregations across masks and top-$k$ computations to enhance the versatility of the supported queries. 
Aggregations across masks serve as a powerful tool for comparing trends of different masks, e.g., studying the difference between model saliency maps and human attention maps~\cite{das2016human}. 
Top-$k$ computations are also widely used. 
For example, Alice might be interested in finding the top-$k$ X-rays whose saliency maps have the least number of high-value pixels in the lung regions.

Efficiently executing the formulated queries is challenging. 
The database of masks is too large to fit in memory, loading all masks from disk is slow and dominates the query execution time, compressing
images does not help due to the overhead of decompression. 
Existing methods and systems do not support these queries efficiently either. 
For instance, using either NumPy or PostgreSQL to load and process the masks, a query that filters masks based on the number of pixels within an ROI and a pixel value range takes more than 30 minutes to complete on ImageNet (see \cref{fig:single-query-performance}).
Although array databases such as SciDB~\cite{scidb2010brown} and TileDB~\cite{papadopoulos2016tiledb} are designed to process multi-dimensional dense arrays, they are not optimized for efficiently searching through large collections of small arrays, as required in these queries (see \cref{fig:single-query-performance}). 
Moreover, existing multi-dimensional indexing techniques also do not provide better execution times because masks are dense arrays.

\system accelerates the aforementioned queries without any loss in query accuracy by introducing a new type of index and an efficient filter-verification query execution framework. 
Both techniques work in tandem to reduce the number of masks that must be loaded from disk during query execution while guaranteeing the correctness of the query result. 
The indexing technique, which we call the Cumulative Histogram Index (CHI), provides bounds on the pixel counts within an ROI and a pixel value range in a mask. 
It is designed to work with arbitrary ROIs (both mask-specific and constant) and pixel value ranges specified by the user at query time. 
These bounds are used during query execution when deciding whether a mask should be loaded from disk and processed while guaranteeing the correctness of the query result. 

\system's query execution employs the idea of pre-filtering. 
Using pre-filtering techniques to avoid expensive computation or disk I/O has been explored and proven to be effective in many other problems, such as accelerating similarity joins~\cite{mann2016anempirical, jiang2014string} and queries that contain ML models~\cite{kang2017noscope, lu2018probabilistic, anderson2018physical, hsieh2018focus} in cases where computing the similarity function or running model inference is expensive during query execution. 
\system's filter-verification execution framework leverages CHI to bypass the loading of the masks that are guaranteed to satisfy or not satisfy the query predicate. 
Only the masks that cannot be filtered out are loaded from disk and processed. 
By doing so, \system overcomes the limitation of existing systems by reducing the number of masks that must be loaded to process a query. 
Moreover, \system includes an incremental indexing approach that avoids potentially high upfront indexing costs and enables it to operate in an online setting. 

In summary, the contributions of this paper are: 

\begin{itemize}[itemsep=1pt, topsep=1pt, leftmargin=10pt]
    \item \revision{We formalize a novel, and broadly applicable, class of queries that retrieve images and their masks from image databases based on the properties of the latter, and further extend the queries to support aggregations across masks and top-$k$ computations (\cref{sec:background}).} 
    \item We develop a novel indexing technique and an efficient filter-verification query execution framework (\cref{sec:system}). 
    \item We implement the algorithms in a prototype system, \system, and demonstrate that it achieves up to two orders of magnitude speedup over existing methods for individual queries and consistently outperforms existing methods on various multi-query workloads that simulate dataset exploration and analysis processes (\cref{sec:evaluation}). 
\end{itemize}

Overall, \system is an important next step toward the seamless and rapid exploration of a dataset based on masks generated by ML models. It is an important component in a toolbox of methods for ML model explainability and debugging.

\section{Queries over Masks}
\label{sec:background}

This section formalizes the queries that \system supports and discusses the challenges associated
with their efficient execution.

\subsection{Data and Query Model}
\label{sec:language}

\noindent\textbf{Data Model.}
An image is a 2D array of pixel values. 
A mask over an image is also a 2D array of pixel values. 
The values in a mask, however, are limited to the range $[0, 1.0)$. 
\cref{fig:query-example} shows an illustrative example of a toy x-ray image and an associated mask. 
The example shows a saliency map in which a higher value means that the pixel is more important to the model's decision. 
We can capture this data model with the following conceptual relational view, 

\small
\begin{verbatim}
MasksDatabaseView (
  mask_id INTEGER PRIMARY KEY,
  image_id INTEGER,
  model_id INTEGER,
  mask_type INTEGER,
  mask REAL[][],
  ... );
\end{verbatim}
\normalsize

where \texttt{mask\_id}, \texttt{image\_id}, and \texttt{model\_id} store the unique identifiers of the mask, image, and model that generate the mask, respectively. 
\texttt{mask\_type} is the identifier of the type of mask (an \texttt{ENUM} type), e.g., saliency map, human attention map, segmentation mask, depth mask, etc. 
The \texttt{mask} column stores the mask itself. 
Each mask is a 2D array of floating points in the range of $[0, 1)$.
Additional columns can store other information, such as ground-truth labels, predicted labels, and image capture times. 
With some abuse of notation, an example tuple in the above view could be $(6, 4, \text{ResNet-50}, \text{SaliencyMap}, [[0.9, 0.5,\ldots], \ldots])$, referring to a saliency map (mask \#6) computed for image \#4 using ResNet-50~\cite{he2016deep}. 
Note that \texttt{mask\_id} does not have a direct relationship with \texttt{image\_id} because an image can have multiple or no masks. 

A region of interest (ROI) is a bounding box, $b$, which can either be user-specified or computed by a query. 
\cref{fig:query-example} shows a user-specified ROI that corresponds to the part of the image with the lungs. 
ROIs are query-dependant, so they are not included in \texttt{MasksDatabaseView} but are computed during query execution. 

\queryExampleFigure

\noindent\textbf{Basic Queries.} 
\system is designed to support queries that specify: 
(1) regions of interest within images (e.g., where the user expects the lungs to be located), 
(2) filter predicates over the pixel values in a mask (e.g., all pixel values above a threshold, indicating importance), and aggregates over those pixels that satisfy the predicates (i.e., count of pixels). 
A query over a mask can be expressed with the following SQL query, where \texttt{opt.} indicates that a clause is optional and concepts like \texttt{roi} will be explained in detail below, 

\small
\begin{verbatim}
SELECT *, CP(mask, roi, (lv, uv)) as val
FROM MasksDatabaseView
WHERE <filter on CP(...)> [AND | OR] ...            -- opt.
ORDER BY val [ASC | DESC] [LIMIT K]                 -- opt.
\end{verbatim}
\normalsize

\noindent\textbf{Region of interest (ROI).}
The ROI, \texttt{roi}, is a bounding box represented by pairs of coordinates that are the upper left and lower right corners of the box.
It can be constant for all masks or different for each mask, e.g., the bounding box of the foreground object in each image.
The ROI is specified by the user at query time or obtained from another table joined with \texttt{MasksDatabaseView}.

\noindent\textbf{\texttt{CP} function.} At the core of the query is the \texttt{CP} function.
It takes in a mask, an ROI, a lower bound (\texttt{lv}), and an upper bound (\texttt{uv}) as input, and returns the number of pixels in the ROI of the mask with values in the range of $[\texttt{lv}, \texttt{uv})$.
\texttt{CP} is formally defined as follows, 

\addtolength{\abovedisplayskip}{-5pt}
\addtolength{\belowdisplayskip}{-5pt}
\small
\setstretch{0.7}
\begin{align*}
\texttt{CP}(mask, roi, (lv, uv)) = \sum_{(x, y) \in roi} \mathbbm{1}_{lv \le mask[x][y] < uv}
\end{align*}
\normalsize
\setstretch{1.0}
\addtolength{\abovedisplayskip}{5pt}
\addtolength{\belowdisplayskip}{5pt}

where $\mathbbm{1}_{\text{condition}}$ is an indicator function that is 1 if the condition is true and 0 otherwise. 
Note that the output of \texttt{CP} is a scalar value and arithmetic operations can be applied to it. 
In our queries, \texttt{CP} is often present in the filter predicate, e.g., $\texttt{CP}(mask, roi, (lv, uv)) > T$, and in the \texttt{ORDER BY} clause, e.g., $\texttt{ORDER BY CP}(mask, roi, (lv, uv)) \texttt{ ASC}$. 
Multiple \texttt{CP} functions can be used in a query, e.g., to specify multiple ROIs, or to compute multiple ratios of pixels in different ranges. 

\noindent\textit{\textbf{Example 1}: Consider Scenario 2 from \cref{sec:introduction}. 
Alice, the scientist, is building a model that takes X-ray images as input and classifies them as COVID-19 vs. non-COVID.  
Her model does not work well once deployed. 
To investigate the problem, Alice wants to verify that the model is focusing its attention on the region in the images that corresponds to the lungs. 
Hence, she writes a query that computes the number of salient (i.e., important, or with value $> 0.85$) pixels within the ROI that corresponds to the lungs, which she specifies manually as a bounding box, \texttt{roi}\footnote{For readability, we specify the ROI as the variable, \texttt{roi}. This would normally be a set of four numbers specifying the coordinates of the bounding box.}. 
She retrieves all the images where the number of salient pixels is less than 10,000 by,}

\small
\begin{verbatim}
SELECT image_id FROM MasksDatabaseView
WHERE CP(mask, roi, (0.85, 1.0)) < 10000;
\end{verbatim}
\normalsize

\noindent\textit{
She can also compute the ratio of the number of salient pixels within the lung region to the total number of salient pixels in the image. She queries the top-25 images with the lowest ratios by, 
}

\small
\begin{verbatim}
SELECT image_id,
  CP(mask, roi, (0.85, 1.0)) / CP(mask, -, (0.85,1.0)) AS r
FROM MasksDatabaseView ORDER BY r ASC LIMIT 25;
\end{verbatim}
\normalsize

\noindent\textbf{Complex Queries.} \system further supports aggregations over pixel counts and pixel counts over aggregated masks. These more complete queries
can be expressed with the following SQL, 

\small
\begin{verbatim}
SELECT [mask_id | image_id | model_id | ...],
 [SCALAR_AGG(CP(mask, roi, (lv, uv)))
 | CP(MASK_AGG(mask), roi, (lv, uv))] as aggregate
FROM MasksDatabaseView
WHERE <filter on CP(...)> [AND | OR] ...            -- opt.
GROUP BY [image_id | model_id | mask_type]          -- opt.
HAVING <filter on aggregate> [AND | OR] ...         -- opt.
ORDER BY aggregate [ASC | DESC] [LIMIT K]           -- opt.
\end{verbatim}
\normalsize

\noindent\textbf{Scalar aggregation.}
The user can aggregate the outputs of \texttt{CP} functions for masks of the same image, model, or mask type, by defining the \texttt{SCALAR\_AGG} function, which aggregates the outputs of \texttt{CP} functions. 
\system supports common functions such as \texttt{SUM}, \texttt{AVG}, \texttt{MIN}, and \texttt{MAX}, e.g., the average of multiple \texttt{CP} functions over masks produced by different models grouped by \texttt{image\_id}. 

\noindent\textbf{Mask aggregation.}
\texttt{MASK\_AGG} is used to aggregate masks themselves.
It is a user-defined function that takes in a list of masks as input and returns a mask: \texttt{MASK\_AGG} $\rightarrow$ \texttt{REAL[][]}. 
An example of \texttt{MASK\_AGG} is \small $\texttt{INTERSECT}(m_1 > 0.8, ..., m_n > 0.8)$\normalsize, i.e., the intersection of $n$ masks after thresholding at 0.8.

\noindent\textit{\textbf{Example 2}: Consider a case where our user in Scenario 2 in \cref{sec:introduction}, Alice, would like to understand if her model focuses on the same parts of the X-ray images as human experts. 
After setting \texttt{roi} to the full mask, she can write the query below, where saliency maps have $\texttt{mask\_type} = 1$ and human attention maps have $\texttt{mask\_type} = 2$, }

\small
\begin{verbatim}
SELECT image_id, CP(INTERSECT(mask > 0.7), roi, (0.7, 1.0)) AS s
FROM MasksDatabaseView WHERE mask_type IN (1, 2)
GROUP BY image_id ORDER BY s DESC LIMIT 10;
\end{verbatim}
\normalsize

\subsection{Challenges} \label{sec:challenges}

Processing the above queries efficiently is challenging.
A baseline approach of loading masks from disk into memory before query processing is extremely slow because it saturates disk read bandwidth.
A single query on ImageNet~\cite{ILSVRC15} takes more than 30 minutes to complete (\cref{fig:single-query-performance}).
Alternatively, storing compressed masks reduces data loaded from disk but moves the bottleneck to decompression, so a single query on ImageNet still takes around 30 minutes.

Existing systems, such as PostgreSQL, have the same bottleneck of loading masks from disk. 
Existing multi-dimensional indexing techniques do not efficiently support our target queries because mask data is dense. 
They require representing each mask's pixel as a point in the space of $(x, y, \text{pixel value})$, where $x$ and $y$ denote pixel coordinates. 
In this space, our query is an orthogonal range query followed by an aggregation by \texttt{mask\_id}. 
The best known algorithm \cite{orr1990chazelle, rangetree1990chazelle}, range trees, has a query time of $O(k + \log^2 n)$ and a preprocessing time of $O(n \log^2 n)$. 
Here, $n$ is \# mask pixels in the dataset, and $k$ is \# pixels in the cuboid defined by \texttt{roi} and \texttt{(lv,uv)}. 
$n$ is extremely large because mask data is dense (e.g., 65 billion for ImageNet), which makes using these indexes infeasible. 
Array databases~\cite{scidb2010brown, papadopoulos2016tiledb} are designed to work with dense arrays, but they are optimized for complex computations over small numbers of large arrays rather than efficiently searching through large numbers of arrays. 
While they can load specific slices within a desired ROI rather than entire arrays, \system avoids loading any pixels at all for a large fraction of masks, as we explain next. 
We discuss related work further in \Cref{sec:related-work}.

\section{\system} \label{sec:system}

\system efficiently executes queries over a database of image masks while guaranteeing the correctness of query results. 
As presented above, the fundamental operations in our target queries involve filtering masks based on pixel values within ROIs, followed by performing optional aggregations, sorting, or top-$k$ computations. 
The key challenge when performing these operations is that the database of masks is too large to fit in memory, and scanning, loading, and processing all masks is slow.

To accelerate such queries, \system introduces a novel type of index, called the Cumulative Histogram Index (CHI) (\cref{sec:chi}), and an efficient filter-verification query execution framework (\cref{sec:filter-verification}). 
The CHI technique indexes each mask by maintaining pixel counts for key combinations of spatial regions and pixel values. 
CHI constructs a compact data structure that enables fast computation of upper and lower bounds on \texttt{CP} functions for arbitrary ROIs and pixel value ranges. 
These bounds are used during query execution to efficiently filter out masks that are either guaranteed to fail the query predicate or guaranteed to satisfy it without loading them from disk. 
The query execution framework comprises two stages: the filter stage and the verification stage. 
During the filter stage, the framework utilizes CHI to compute bounds on \texttt{CP} functions to filter out the masks without loading them from disk. 
Then, during the verification stage, the framework verifies the remaining masks by loading them from disk and applying the full predicate. 
This framework guarantees the correctness of the query results and overcomes the bottleneck of query execution by significantly reducing the number of masks that must be loaded from disk. 

\subsection{Cumulative Histogram Index (CHI)} \label{sec:chi}

The key goals of CHI are to: 
(\textbf{G1}) support arbitrary query parameters $lv$ and $uv$ that specify the range of pixel values, which are unknown to \system ahead of time, and 
(\textbf{G2}) support arbitrary regions of interest, $roi$, and allow mask-specific $roi$s in a single query. 
The $roi$s are also unknown ahead of time because the user can specify $roi$s arbitrarily at query time.

\noindent\textbf{Key Idea.} 
\system achieves both goals by building CHI to maintain pixel counts for different combinations of spatial locations and pixel values. 
Conceptually, \system builds an index on the search key $(mask\_id, roi, \text{pixel value})$. 
For each search key, CHI holds the count of pixels that satisfy the condition. 
Given $mask\_id$, $roi$, and a range of pixel values specified by $(lv, uv)$, the index supports queries that return the number of pixels in the $roi$ of the mask with values in the range $(lv, uv)$, i.e., $\texttt{CP}(mask, roi, (lv, uv))$. 

Building an index on every possible combination of $(mask\_id, roi, \text{pixel value})$ is infeasible both in terms of space and time complexity because the number of possible $roi$s for each mask is quadratic in the number of pixels in the mask, let alone the number of masks and the number of pixel values. 

Instead, CHI builds a data structure that efficiently provides upper and lower bounds on predicates, rather than exact values. 
This approach leads to a small-size index while still effectively pruning masks that are either guaranteed to fail the predicate or guaranteed to satisfy it. 
Only a small fraction of masks must then be loaded from disk and processed in full to verify the predicate.

\noindent\textbf{CHI Details.} CHI leverages two key ideas: discretization and cumulative counts. 
Discretization reduces the total amount of information in the index, while cumulative counts yield highly efficient lookups. 
We explain both here.

To build a small-sized index, \system partitions masks into disjoint regions and discretizes pixel values into disjoint intervals. 
It then builds an index on the combinations of $(mask\_id, \text{region}, \text{pixel value interval})$. 
\revision{For the spatial dimensions, \system partitions each mask into a grid of cells, each of which is $w_c$ by $h_c$ pixels in size.} 
For the pixel value dimension, \system discretizes the values into $b$ buckets (bins). 
\system could use either equi-width or equi-depth buckets. 
Our current prototype uses equi-width buckets.

After discretization, there are several options for implementing the index. 
A straightforward option is to build an index on the search key $(mask\_id, cx\_id, cy\_id, bin\_id)$, where $cx\_id$, $cy\_id$, and $bin\_id$ identify the coordinates of each unique combination of grid and pixel-value range (e.g., $cx\_id$ of 3 corresponds to the grid cell that starts at pixel $w_c * 3$, similarly for $cy\_id$ and $bin\_id$). 
For each such key, the index could store the number of pixels in the mask whose coordinates are in the cell identified by $(cx\_id, cy\_id)$ and with values in the range $[p_{min} + bin\_id\cdot\Delta, p_{min} + (bin\_id + 1)\cdot\Delta)$, where $p_{min}$ is the lowest pixel value across all masks and $\Delta$ is the width of each bucket. 
This option would require \system to identify all the cells that intersect with $roi$ and all the bins that intersect with $(lv, uv)$ and perform our query execution technique (discussed in \cref{sec:filter-verification}) on the pixel counts of these cells and bins. 
A more efficient approach, which we adopt, is to build an index on the search key $(mask\_id, cx\_id, cy\_id, bin\_id)$, but, for each key, store the reverse cumulative sum of pixel counts in the mask with values in the range $[p_{min} + bin\_id\cdot\Delta, p_{max}]$ and coordinates in the region of $((1, 1), (cx\_id \cdot w_c, cy\_id \cdot h_c))$. 
This index is denoted with $H(mask\_id, cx\_id, cy\_id, bin\_id)$. 
We will also use $H(mask\_id, cx\_id, cy\_id)$ to denote the array of cumulative sums for all bins, i.e., $H(mask\_id, cx\_id, cy\_id)[bin\_id] = H(mask\_id, cx\_id, cy\_id, bin\_id)$. 
Recall that $mask\_id$ uniquely identifies $mask$. 
The index can be formally expressed as, 

\setstretch{0.7}
\small
\begin{align} \label{eq:histogram-index}
\begin{split}
  H(mask\_id&, cx\_id, cy\_id, bin\_id) \\
  = \texttt{CP}(&mask, ((1, 1), (cx\_id \cdot w_c, cy\_id \cdot h_c)), \\
                    &(p_{min} + bin\_id\cdot\Delta, p_{max}))
\end{split}
\end{align}
\normalsize
\setstretch{1.0}

\chiIllustrationFigure

\noindent\textit{\textbf{Example}: 
In \cref{fig:chi-illustration}, \system builds CHI for an example mask, $M$, with $w_c = 2$ , $h_c = 2$, and $b = 2$. 
Hence, each cell, $(x_c, y_c)$, highlighted in light blue marks the corner of a discretized region. 
With $b=2$, the pixel value range is discretized into $b$ bins, $[0, 0.5)$ and $[0.5, 1)$. 
\system builds $H(M, x_c / w_c, y_c / h_c)$ for each of the corner cells. 
For example, for cell $(2,2)$, we have $H(M, 1, 1)[0] = 4$ (all four pixels are in $(p_{min}, p_{max})$) and $H(M, 1, 1)[1] = 0$ (no pixels are in the $0.5$ to $p_{max}$ range). 
For cell, $(4,4)$, $H(M,2,2) = [16,3]$.
}

In essence, $H(mask\_id, cx\_id, cy\_id, bin\_id)$ stores a cumulative sum of pixel counts, considering both spatial and pixel value dimensions. 
Storing cumulative sums offers greater efficiency compared to storing raw values, as it enables rapid evaluation of pixel counts within a specific range, in terms of both spatial and pixel value dimensions, by only performing simple arithmetic operations without having to access all the bins within the desired pixel value range for all the cells in the desired spatial region. 
To illustrate this, we first introduce the concept of \textit{available regions}. 

\addtolength{\abovedisplayskip}{-10pt} %
\addtolength{\belowdisplayskip}{-10pt} %
\begin{definition} \label{def:available-region}
  Let $X_c$ denote $\{x_c | x_c \in [w_c, 2w_c, 3w_c \dots, w]\}$ and $Y_c$ denote $\{y_c | y_c \in [h_c, 2h_c, 3h_c, \dots, h]\}$. A region $((x_1, y_1), (x_2, y_2))$ is \textit{available} in the CHI of a mask if $(x_2, y_2) \in X_c \times Y_c$ and $(x_1 - 1, y_1 - 1) \in (X_c \cup \{0\}) \times (Y_c \cup \{0\})$.
\end{definition}
\addtolength{\abovedisplayskip}{10pt} %
\addtolength{\belowdisplayskip}{10pt} %

\noindent\textit{\textbf{Example:} Available regions in \cref{fig:chi-illustration} are bounding boxes that start from the bottom-right corner of a blue cell\footnote{$(0, 0)$, not shown in the figure, is considered as a blue cell as well.} and end at the bottom-right corner of a blue cell, e.g., $((3, 3), (4, 6))$ is an available region, highlighted with a dark green bounding box; $((4, 4), (5, 5))$ is not an available region, highlighted with an orange bounding box.}

Pixel counts within \textit{available regions} are used to compute bounds on \texttt{CP} functions for arbitrary ROIs and pixel value ranges during query execution (\cref{sec:filter-verification}). 
Before we get to these bounds, we first explain how to compute pixel counts within an \textit{available region} with pixel values within the range of two bin boundaries, \system performs the following steps: 
(1) compute the reverse cumulative sums for the specified region using the index values; 
(2) calculate pixel counts between the two bin boundaries by subtracting the relevant cumulative sums. 
The details are explained below. 

Let $C(mask\_id, r)$ denote the histogram of the reverse cumulative pixel counts of region $r$ in mask $mask\_id$, where $C(mask\_id, r)[i] = \texttt{CP}(mask, r, (p_{min} + i\Delta, p_{max}))$. 
\system can compute $C(mask\_id, ((x_1, y_1), (x_2, y_2)))$ for any \textit{available region} $((x_1, y_1), (x_2, y_2))$. 
Let $M$ denote $mask\_id$ for brevity, we have, 

\setstretch{0.7}
\small
\begin{align} \label{eq:subregion-histogram}
  \begin{split}
      &C(M, ((x_1, y_1), (x_2, y_2))) \\
      = &\text{ }H(M, x_2 / w_c, y_2 / h_c) - H(M, (x_1 - 1) / w_c, y_2 / h_c) \\
      - &\text{ }H(M, x_2 / w_c, (y_1 - 1) / h_c) + H(M, (x_1 - 1) / w_c, (y_1 - 1) / h_c) \\
  \end{split}
\end{align}
\normalsize
\setstretch{1.0}

\additiveFunctionIllustrationFigure

\noindent where $-$ and $+$ are element-wise subtraction and addition, respectively, for two arrays of the same size. 
\cref{eq:subregion-histogram} holds because $C(mask\_id, region)$ is a (finitely)-additive function over disjoint spatial partitions since each bin of $C(mask\_id, region)$ is a \texttt{CP} function which is (finitely)-additive. 
\cref{fig:additive-function-illustration} shows an illustrative example of this additive property. 
Note that for any $mask\_id$ and $roi$, $C(mask\_id, roi)[\lceil p_{max} / \Delta \rceil]$ is always $0$ for notation simplicity. 

\noindent\textit{\textbf{Example}: \cref{fig:chi-illustration} shows how $C(M, ((3, 3), (4, 6)))$ is computed.}

After \system computes the reverse cumulative sums of pixel counts, $C$, for a region $r$, the pixel counts between any two bin boundaries (for pixel value discretization) can be obtained by subtracting the cumulative sums of the two bins.

Given a predicate $\texttt{CP}(mask, roi, (lv, uv)) > T$, \system uses CHI to check whether the predicate is satisfied. 
At a high level, \system identifies \textit{available regions}, $r_1$ and $r_2$, in the CHI of the mask, such that $r_1$ is the smallest region that covers $roi$ and $r_2$ is the largest region that is covered by $roi$. 
Then, \system computes $C(mask, r_1)$ and $C(mask, r_2)$ using \cref{eq:subregion-histogram} and uses them to compute the lower and upper bounds of $\texttt{CP}(mask, roi, (lv, uv))$. 
Finally, \system checks whether $mask$ is guaranteed to satisfy or guaranteed to fail the predicate by comparing the lower and upper bounds with $T$. 
The details are further explained in \cref{sec:filter-verification}. 

Since $mask\_id$, $cx\_id$, $cy\_id$, and $bin\_id$ are all integers, rather than building a B-tree index or a hash index over the keys, we create an optimized index structure using an array where $mask\_id$, $cx\_id$, $cy\_id$, and $bin\_id$ act as offsets for lookups in the array. 
We call this structure the Cumulative Histogram Index (CHI) and $H(mask\_id)$ the CHI of mask $mask\_id$. 
There are several advantages of this optimized index structure. 
First, it enables \system to only store the values of CHI and avoid the cost of storing the keys of CHI and the overhead of building a B-tree or hash index. 
Second, for any lookup key, the lookup latency is of constant complexity and avoids pointer chasing which is common in other index structures. 

\revision{
The time complexity for computing CHI for $N$ masks of size $w \times h$ is $O(N \cdot w \cdot h)$, and this cost is amortized over time with the incremental indexing technique described in \cref{sec:incremental-indexing}.
} 
The number of CHI that \system builds for $N$ masks is $N \cdot w \cdot h / (w_c \cdot h_c)$. 
Each CHI has $b$ bins, thus taking $4 \cdot b$ bytes. 
Hence, the set of CHI for $N$ masks takes $4 \cdot b \cdot N \cdot w \cdot h / (w_c \cdot h_c)$ bytes in space. 
With a reasonable configuration of $b$, $w_c$, and $h_c$, CHI can be held in memory for a moderately-sized dataset, and \system can achieve good query performance with it (see \cref{sec:eval-single-query-workload-motivation}).

\subsection{Filter-Verification Query Execution} \label{sec:filter-verification}

Without loss of generality, in this section, we will show how \system accelerates the execution of a one-sided filter predicate $\texttt{CP}(mask, roi, (lv, uv)) > T$, denoted with $P$, as multiple one-sided filter predicates can be combined to form a complex predicate. 
In \cref{sec:generic-predicates}, we will show that our technique applies to accelerating predicates that are in the form of $\texttt{CP}(...) < T$ or involve multiple different \texttt{CP} functions, e.g., $\texttt{CP(...) < CP(...)}$. 
Aggregations and top-$k$ queries are discussed in \cref{sec:aggregation} and \cref{sec:top-k-queries}, respectively.

\system takes as input a filter predicate $P$, and its goal is to find and return the $mask\_id$s of the masks that satisfy $P$. 
At a high level, \system executes the following workflow:

\begin{itemize}[itemsep=1pt, topsep=1pt, leftmargin=12pt]
    \item \textbf{Filter stage}: filter out the masks that \textit{are guaranteed to fail} the predicate $P$, and add the masks that \textit{are guaranteed to satisfy} the predicate $P$ directly to the result set, before loading them from disk to memory. 
    \item \textbf{Verification stage}: load the remaining unfiltered masks from disk to memory and verify them by applying predicate $P$. If a mask satisfies $P$, add it to the result set.
\end{itemize}

\revision{It is worth noting that \system guarantees the correctness of the query results because it only prunes the masks that are guaranteed to fail $P$ and adds the masks that are guaranteed to satisfy $P$ directly to the result set; it subsequently verifies any uncertain masks to ensure result correctness.}

\subsubsection{Filter Stage} \label{sec:filter-stage}

At a high level, the algorithm works as follows, for each mask, \system uses the CHI of the mask to compute bounds of $\texttt{CP}(mask, roi, (lv, uv))$ and it then uses the bounds to determine whether the mask will satisfy $P$ or not. 
In this manner, \system reduces the number of masks loaded from disk during the verification stage (\cref{sec:verification-stage}) by pruning the masks that are guaranteed to fail $P$ and adding the masks that are guaranteed to satisfy $P$ directly to the result set $R$. 
Deriving the bounds of $\texttt{CP}(mask, roi, (lv, uv))$ is challenging because $roi$ and $(lv, uv)$ can be arbitrary and not known in advance. 
\system addresses this challenge by leveraging the CHI of masks and the (finitely)-additive property of CHI to derive the bounds for arbitrary $roi$ and $(lv, uv)$.

\noindent\textbf{Notation.} 
$P$ denotes $\texttt{CP}(mask, roi, (lv, uv)) > T$. 
$mask$ is uniquely identified by $mask\_id$. 
$\theta$ denotes the actual value of $\texttt{CP}(mask, roi, (lv, uv))$. 
$\bar{\theta}$ and $\ubar{\theta}$ denote the upper bound and the lower bound on $\theta$ computed by \system, respectively. 
$C(mask\_id, r)$ denotes the histogram of reverse cumulative pixel counts of the pixel value bins of region $r$ in mask $mask\_id$, 
where $C(mask\_id, r)[i] = \texttt{CP}(mask, r, (p_{min} + i\Delta, p_{max}))$.

When a session of \system starts, the CHI of each mask is loaded from disk to memory and will be held in memory for the duration of the system run time. 
In cases where CHI cannot be held in memory, \system loads the CHI of a mask from disk on demand during query execution. 
Note that the size of the CHI of a mask is much smaller than the size of the mask itself, and therefore, even if the CHI of a mask is on disk, computing the bounds is much less expensive than loading the masks from disk to memory and evaluating the predicate $P$ on them. 

Given a predicate $P$, \system processes each mask targeted by the filter predicate in parallel. 
For each $mask$ uniquely identified by $mask\_id$, \system proceeds as follows:

\vspace{0.25em}
\noindent\textbf{Step 1: Compute $\bar{\theta}$ and $\ubar{\theta}$.} 
In this step, \system computes $\bar{\theta}$ and $\ubar{\theta}$ by using the CHI of $mask\_id$. 
\system uses two approaches to computing two upper bounds, $\bar{\theta}_1$ and $\bar{\theta}_2$, on $\theta$, and uses the smaller one as $\bar{\theta}$. 
The two approaches are effective in yielding bounds in different scenarios (details below). 

Approach 1 first identifies the smallest \textit{available region} (\cref{def:available-region}) in the CHI that covers $roi$ of $mask\_id$. 
We denote this region with $\lowoverline{roi}$. 
Then, $C(mask\_id, \lowoverline{roi})$ can be computed by CHI using \cref{eq:subregion-histogram}. 
Finally, $\bar{\theta}_1$ is computed as, 

\setstretch{0.7}
\small
\begin{align} \label{eq:upper-bound-1}
  \begin{split}
      \bar{\theta}_1 = C(mask\_id, \lowoverline{roi})[\lfloor lv / \Delta \rfloor] - C(mask\_id, \lowoverline{roi})[\lceil uv / \Delta \rceil]
\end{split}
\end{align}
\normalsize
\setstretch{1.0}

\noindent where $\lfloor x \rfloor$ and $\lceil x \rceil$ denote the floor and ceiling of $x$, respectively. 

Approach 2 first identifies the largest \textit{available region} (\cref{def:available-region}) covered by $roi$ in the CHI for each mask. 
We denote this region with $\underline{roi}$. 
Then, $C(mask\_id, \underline{roi})$ can be computed using \cref{eq:subregion-histogram}. 
Finally, $\bar{\theta}_2$ is computed as, 

\setstretch{0.7}
\small
\begin{align} \label{eq:upper-bound-2}
\begin{split}
    \bar{\theta}_2 = &\text{ }C(mask\_id, \underline{roi})[\lfloor lv / \Delta \rfloor] - C(mask\_id, \underline{roi})[\lceil uv / \Delta \rceil] \\
     & + |roi| - |\underline{roi}|
\end{split}
\end{align}
\normalsize
\setstretch{1.0}

\noindent where $|\cdot|$ denotes the area of a region. 

Finally, $\bar{\theta}$ is computed by taking the minimum of $\bar{\theta}_1$ and $\bar{\theta}_2$. 
To show $\bar{\theta}$ is an upper bound of $\theta$, we first show the following inequality. 
Because $(\lfloor lv / \Delta \rfloor * \Delta, \lceil uv / \Delta \rceil * \Delta)$ is a superset of $(lv, uv)$, for any $mask\_id$ and $roi$, we have,

\setstretch{0.7}
\small
\begin{align} \label{lemma:bin-overlap-bound}
  C(mask\_id, roi)[\lfloor lv / \Delta \rfloor] - C(mask\_id, roi)[\lceil uv / \Delta \rceil] \geq \theta
\end{align}
\normalsize
\setstretch{1.0}

We now show the following theorem. 
\vspace{-0.25em}
\begin{theorem}
    \label{thm:upper-bound}
    $\bar{\theta}$ is an upper bound of $\theta$.
\end{theorem}
\vspace{-0.25em}
We prove the theorem by showing both $\bar{\theta}_1 \geq \theta$ and $\bar{\theta}_2 \geq \theta$. 
For conciseness, we omit $mask\_id$ in $C(mask\_id, ...)$ and omit $mask$ in $\texttt{CP}(mask, ...)$ when clear from context, i.e., $C(Q)$ denotes $C(mask\_id, Q)$ and $\texttt{CP}(Q, (lv, uv))$ denotes $\texttt{CP}(mask, Q, (lv, uv))$. 
We also use $\texttt{CP}(Q \setminus W, (lv, uv))$ to denote the count of pixels in spatial region $Q \setminus W$ with pixel values in $(lv, uv)$. 
\vspace{-0.35em}
\begin{proof}

We first show $\bar{\theta}_1 \geq \theta$. 

\setstretch{0.7}
\small
\begin{align}
\bar{\theta}_1 = & \text{ } C(\lowoverline{roi})[\lfloor lv / \Delta \rfloor] - C(\lowoverline{roi})[\lceil uv / \Delta \rceil] \\
               \geq & \text{ } \texttt{CP}(\lowoverline{roi}, (lv, uv))\label{eq:upper-bound-1-pixel-value-partition} \\ 
               = & \text{ } \texttt{CP}(roi, (lv, uv)) + \texttt{CP}(\lowoverline{roi} \setminus roi, (lv, uv))\label{eq:upper-bound-1-spatial-partition} \\
               \geq & \text{ } \texttt{CP}(roi, (lv, uv)) = \theta \label{eq:upper-bound-1-geq-theta}
\end{align}
\normalsize
\setstretch{1.0}

where Inequality~(\ref{eq:upper-bound-1-pixel-value-partition}) follows from \cref{lemma:bin-overlap-bound} and \cref{eq:upper-bound-1-spatial-partition} follows from \texttt{CP} is an additive function over disjoint spatial regions. 

Let $L$ denote $(\lfloor lv / \Delta \rfloor * \Delta, \lceil uv / \Delta \rceil * \Delta)$. 
We now show $\bar{\theta}_2 \geq \theta$. 

\setstretch{0.7}
\small
\begin{align}
\theta = & \text{ } \texttt{CP}(roi, (lv, uv)) \\
\leq & \text{ } \texttt{CP}(roi, L) \label{eq:upper-bound-2-pixel-value-partition} \\ 
       = & \text{ } \texttt{CP}(\underline{roi}, L) + \texttt{CP}(roi \setminus \underline{roi}, L) \label{eq:upper-bound-2-spatial-partition}\\
       \leq & \text{ } \texttt{CP}(\underline{roi}, L) + |roi| - |\underline{roi}| \label{eq:upper-bound-2-pixel-count-area-bound} \\ 
       = & \text{ }C(\underline{roi})[\lfloor lv / \Delta \rfloor] - C(\underline{roi})[\lceil uv / \Delta \rceil] + |roi| - |\underline{roi}| \label{eq:upper-bound-2-ct-px-c} = \text{ } \bar{\theta}_2 
\end{align}
\normalsize
\setstretch{1.0}

where \cref{eq:upper-bound-2-spatial-partition} follows from the fact that \texttt{CP} is an additive function over disjoint spatial regions. 
Inequality~(\ref{eq:upper-bound-2-pixel-count-area-bound}) is because the count of pixels in any region with pixel values in any range is bounded by the total number of pixels in the region. \qedhere
\end{proof}

\upperBoundIllustrationFigure

\noindent\textit{
\textbf{Example}: The two approaches are illustrated with an example mask in \cref{fig:upper-bound-illustration}. 
Mask data is the same as in \cref{fig:chi-illustration}. 
The first approach identifies $\lowoverline{roi}$, which is $((3, 3), (6, 6))$, and $C(M, \lowoverline{roi})$ is computed using \cref{eq:subregion-histogram}. 
Then, $\bar{\theta}_1$ is computed using \cref{eq:upper-bound-1}, i.e., $C(M, \lowoverline{roi})[1] - C(M, \lowoverline{roi})[2] = 8 - 0 = 8$. 
The second approach identifies $\underline{roi}$, which is $((3, 3), (4, 4))$, and $C(M, \underline{roi})$ is computed using \cref{eq:subregion-histogram}. 
Then, $\bar{\theta}_2$ is computed using \cref{eq:upper-bound-2}, i.e., $C(M, \underline{roi})[1] - C(M, \underline{roi})[2] + |roi| - |\underline{roi}| = 2 - 0 + 9 - 4 = 7$.
}

The two approaches are effective in yielding bounds in different scenarios. 
Intuitively, the first approach is more effective when $roi$ and $\lowoverline{roi}$ are close to each other, which would result in a small difference between $\bar{\theta}_1$ and $\theta$. 
The second approach is more effective when $roi$ and $\underline{roi}$ are close to each other. 

The lower bound, $\ubar{\theta}$, can be computed similarly following the two approaches. 
Due to space constraints, we omit the details here.

\vspace{0.25em}
\noindent\textbf{Step 2: Determine the relationship between $\bar{\theta}$ and $\ubar{\theta}$ and $T$.} 
In this step, \system determines whether the predicate $P$ is satisfied by the mask based on the relationship between $\bar{\theta}$ and $\ubar{\theta}$ and $T$. 
There are three cases: 
\begin{itemize}[itemsep=1pt, topsep=1pt, leftmargin=12pt]
  \item \textit{Case 1:} $\bar{\theta} \leq T$. The mask is pruned because it is impossible for the mask to satisfy the predicate $P$. 
  \item \textit{Case 2:} $\ubar{\theta} > T$. The mask is directly added to the result set $R$ because the mask is guaranteed to satisfy the predicate $P$. 
  \item \textit{Case 3:} $\ubar{\theta} \leq T < \bar{\theta}$. The mask is added to the candidate mask set $S$ since it needs to be verified against $P$ in the verification stage.
\end{itemize}

\subsubsection{Verification Stage} \label{sec:verification-stage}

The verification stage aims to verify each candidate mask in $S$ that was neither pruned nor directly added to the result set. 
By loading it from disk and computing the actual value of $\theta$, and then evaluating the predicate $P$, \system determines whether the mask satisfies the predicate $P$. 
If the mask satisfies the predicate $P$, it is added to the result set $R$.

\subsection{Generic Predicates} \label{sec:generic-predicates}
So far, we have described how \system can efficiently process predicates in the form of $\texttt{CP}(mask, roi, (lv, uv)) > T$. 
Supporting predicates in the form of $\texttt{CP}(mask, roi, (lv, uv)) < T$ is similar to the previous case. 
The only difference is that in Step 2 of the filter stage, \system directly adds the mask to the result set $R$ if $\bar{\theta} < T$ and prunes the mask if $\ubar{\theta} \geq T$.

\system also supports generic predicates that involve multiple \texttt{CP} functions, i.e., $\texttt{CP}_1(...) \operatorname{op_1} \texttt{CP}_2(...) \cdots \operatorname{op_{n-1}} \texttt{CP}_n(...) > T$. 
Let $F = \texttt{CP}_1(...) \operatorname{op_1} \texttt{CP}_2(...) \cdots \operatorname{op_{n-1}} \texttt{CP}_n(...)$. 
\system uses the lower and upper bounds of every \texttt{CP} function to derive the lower and upper bounds of $F$ and use the bounds to efficiently prune the masks that are guaranteed to fail the predicate or guaranteed to satisfy it in the filter stage described in \cref{sec:filter-stage}, as long as $F$ is monotonic with respect to each $\texttt{CP}_i$ function. 
Common operators that make $F$ monotonic include $+, -, \times$.

\subsection{Aggregation} \label{sec:aggregation}

\system supports queries that contain scalar aggregates on \texttt{CP} functions or on the \texttt{CP} function over mask aggregations, as described in \cref{sec:background}. 
For filter predicates on scalar aggregates, e.g., $\texttt{SUM}(\texttt{CP}(mask, roi, (lv, uv))) > T$ group by $image\_id$, \system uses the same approach as in \cref{sec:generic-predicates} to efficiently filter out groups of masks associated with the same $image\_id$ that are guaranteed to fail the predicate or guaranteed to satisfy it, since common scalar aggregate functions (\texttt{SUM}, \texttt{AVG}, and etc.) are monotonic with respect to the \texttt{CP} function. 
For filter predicates on mask aggregations, e.g., $\texttt{CP}(\texttt{MASK\_AGG}(mask), roi, (lv, uv)) > T$, \system treats the aggregated masks as new masks and uses the same approach described in \cref{sec:filter-verification} to process the query. 
\revision{
The index for the aggregated masks is either built ahead of time or incrementally built (\cref{sec:incremental-indexing}), which is a limitation of the current prototype. 
However, when the mask aggregation is monotonic, e.g., weighted sum, \system can be easily extended to support efficient filtering of the aggregated masks using indexes built for the individual masks.
}

\subsection{Top-K} \label{sec:top-k-queries}

To answer top-k queries, \system follows a similar idea as described in \cref{sec:filter-verification}, but it intertwines the filter and verification stages to maintain the current top-$k$ result. Without loss of generality, let's consider the case of a top-K query seeking
the masks with the highest values of the \texttt{CP} function.
The set of top-$k$ masks can be defined as a set, $R$, of $k$ masks. 
$R$ is initially empty and is conceptually built incrementally as the query is executed by identifying and adding to $R$ the next mask, $mask$ (associated with its $\texttt{CP}(mask, roi, (lv, uv))$ value), that satisfies the following condition,

\setstretch{0.7}
\small
\begin{align}
  \label{eq:top-k}
  \texttt{CP}(mask, roi, (lv, uv)) > \min_{mask' \in R}{\texttt{CP}(mask', roi, (lv, uv))}
\end{align}
\normalsize
\setstretch{1.0}

\system sequentially processes the masks.  
For each mask, it computes the upper bound $\bar{\theta}$ and compares $\bar{\theta}$ with the \texttt{CP} values of the current $R$. 
If $\bar{\theta} \leq \min_{mask' \in R}{\texttt{CP}(mask', roi, (lv, uv))}$, the mask is pruned because it is impossible for the mask to be in the top-$k$ result; 
otherwise, \system loads the mask from disk and computes the actual value of $\texttt{CP}(mask, roi, (lv, uv))$. 
It then updates $R$ by adding the mask to $R$ if it satisfies \cref{eq:top-k}.

\queryBasedOnMotivationTable

\subsection{Incremental Indexing} \label{sec:incremental-indexing}
As we show in \cref{sec:eval-single-query-workload-motivation} and \cref{sec:eval-single-query-workload-types}, the vanilla \system system described so far achieves a significant query time improvement over the baselines with a small index size. 
The approach that vanilla \system uses, however, incurs a potentially high overhead during preprocessing to build the index. 
Before processing any query, the vanilla \system approach must build the CHI for every mask in the database, which could lead to a long wait time for the user to get the first result. 

To address this challenge, we propose building CHI incrementally as queries are executed so that only the masks that are necessary for the current query are indexed. 
Every time the user issues a query, as \system sequentially processes each mask as described in \cref{sec:filter-verification}, it checks if the CHI of the mask is already built. 
If so, \system directly proceeds as described in \cref{sec:filter-verification}. 
If not, \system executes the query by loading the masks from disk and evaluating whether they satisfy the query predicates. 
For each mask loaded from disk, \system then builds the CHI for the mask and keeps it in memory for future queries in the same session. 
When a \system session ends, the CHI for all the masks in the session is persisted to disk for future sessions. 
With this approach, the cost of building the CHI of a mask is incurred once the first time the mask is loaded from disk, and only if the mask is necessary for a query. 
In \cref{sec:eval-multi-query-workload}, we show that \system with such incremental indexing amortizes the cost of indexing quickly and significantly outperforms other baseline methods on multi-query workloads.
\vspace{-0.5em}
\section{Evaluation} \label{sec:evaluation}

\subsection{Experimental Setup}

\noindent\textbf{Implementation.} 
\system is written in Python as a library. 

\noindent\textbf{Dataset.}
We evaluate \system on two pairs of datasets and models.
The first pair of dataset and model, called \textit{WILDS}, is from~\cite{wilds2020}.
\textit{WILDS} contains \num{22275} images from the in-distribution (ID) and out-of-distribution (OOD) validation sets of the iWildCam dataset~\cite{wilds2020}. 
\revision{
For each image, we use GradCAM~\cite{gradcam2017} to generate two saliency maps for two different ResNet-50~\cite{he2016deep} models obtained from~\cite{wilds2020}. 
Each saliency map is $448 \times 448$ pixels because \textit{WILDS} images (of varying sizes) are resized to be $448 \times 448$ before being fed into the ResNet-50 models from~\cite{wilds2020}. 
The second, called \textit{ImageNet}, contains \num{1331167} images from the ImageNet dataset~\cite{ILSVRC15}. 
We also use GradCAM~\cite{gradcam2017} to generate two saliency maps for two different ResNet-50~\cite{he2016deep} models for each image, and use them as the masks for \textit{ImageNet}. 
Each mask in \textit{ImageNet} is $224 \times 224$ pixels because the models expect images of this size as input~\cite{he2016deep}.
}
These two pairs of models and datasets complement each other in terms of the number of images (and masks) and the size of the masks. 

\noindent\textbf{\system configuration.} Unless otherwise specified, we set $b$ (the number of buckets for pixel value discretization) to $16$ for both \textit{WILDS} and \textit{ImageNet};
then we set $w_c = h_c = 64$ (the cell size for spatial partitioning) for \textit{WILDS} and $w_c = h_c = 28$ for \textit{ImageNet}, 
such that the uncompressed index sizes for both datasets are around $5\%$ of the compressed dataset sizes: the uncompressed index size is 6.5 GB for \textit{ImageNet} and 88 MB for \textit{WILDS}. 
The effect of more granular indexes is discussed in \cref{sec:eval-query-time-analysis}.

\noindent\textbf{Baselines.}
As discussed in \cref{sec:introduction} and \cref{sec:related-work}, there is a lack of system support for the efficient processing of our targeted queries. 
To the best of our knowledge, no existing system reduces the work required, i.e., loading the masks from disk and computing the \texttt{CP} function values for them, to process a query. 
Thus, we compare \system to the following three baselines: 
(1) PostgreSQL 10. 
The masks are stored as 2D arrays of floating point numbers in a column as described in \cref{sec:background}. 
The \texttt{CP} function is implemented as a user-defined function (UDF) written in C and compiled into a dynamically shared library.
It is loaded by the PostgreSQL server when the \texttt{CP} function is called. 
(2) TileDB 2.17.1~\cite{papadopoulos2016tiledb} with TileDB-Py 0.23.1. 
The masks are stored as a 3D array of floating point numbers, with the first dimension being the mask ID, and the second and third dimensions being the height and width of the mask, respectively. 
The tile sizes for \textit{WILDS} and \textit{ImageNet} are set to $448 \times 448$ and $224 \times 224$, respectively because we found that these tile sizes provide the best performance for TileDB as compared to smaller tile sizes. 
(3) NumPy 1.21.6. 
The masks are stored as NumPy arrays on disk. 
The \texttt{CP} function is implemented in Python and uses NumPy array functions to ensure vectorized computation.

\noindent\textbf{Machine configuration.}
All experiments were run on an AWS EC2 p3.2xlarge instance, which has an Intel Xeon E5-2686 v4 processor with 8 vCPUs and 61 GiB of memory, an NVIDIA Tesla V100 GPU with 16 GiB of memory, and EBS gp3 volumes provisioned with 3000 IOPS and 125 MiB/s throughput for disk storage. 
\revision{
We evaluate \system on a single-node setup because most data scientists today work with a single machine~\cite{datascientist2021}. 
Even in a multi-node setup, \system still reduces the number of masks loaded from disk (or over the network) and processed to answer a query, which is the dominant cost of query execution. 
The GPU was used to compute the masks. 
All evaluated methods were using all vCPUs. }

\vspace{-0.5em}
\subsection{Individual Query Performance} \label{sec:eval-single-query-workload-motivation}

\singleQueryBasedOnMotivationFigure

We first evaluate the performance of \system on 5 individual queries motivated by the use cases in \cref{sec:introduction} and \cref{sec:background}: 

\begin{itemize}[itemsep=1pt, topsep=1pt, leftmargin=10pt]
    \item Q1 (Filter, Scenario 2 in~\cref{sec:introduction}): mask selection with a filter predicate on \texttt{CP} with a constant $roi$ across all masks. 
    \item Q2 (Filter, a variant of Q1): mask selection with a filter predicate on \texttt{CP} with different $roi$s for different masks. 
    \item Q3 (Top-K, a variant of Example 1 in~\cref{sec:background}): top-$k$ mask selection, ranked by \texttt{CP} with a constant $roi$ across all masks. 
    \item Q4 (Aggregation, a variant of Example 2 in~\cref{sec:background}): image selection with an aggregation over the \texttt{CP} values of masks associated with different models, with a filter predicate on the aggregated values. 
    \item Q5 (Mask Aggregation, Example 2 in~\cref{sec:background}): image selection with a filter predicate on the \texttt{CP} value of the aggregated mask computed from the masks associated with different models. 
\end{itemize}

The specific parameters for each query are shown in \cref{tab:query-based-on-motivation}. 
\revision{$k$ is set to $25$ for top-$k$ queries because it is a reasonable number of masks to examine for a scientist.} 
When $roi$ is set to object, the $roi$ is the bounding box of the foreground object in the image generated by YOLOv5 \cite{yolov5}. 
We build the CHI for all masks prior to executing the benchmark queries and clear the OS page cache before each query execution. 
The median execution time of 5 runs for each query is shown in \cref{fig:single-query-performance}. 
In addition, \cref{tab:masks-loaded} displays the number of masks loaded from disk by each system during query execution.

As \cref{fig:single-query-performance} shows, on \textit{WILDS}, it takes PostgreSQL, TileDB, and NumPy around 2 minutes to answer each query; on \textit{ImageNet}, it takes them more than 30 minutes to answer each query. 
Profiling these queries showed that mask-loading from disk dominates the query execution time. 
All baseline methods suffer from the same performance bottleneck: they all load all masks from disk and process them to generate the query results. 
Q4 notably takes more time than the other queries. 
This is because it demands the loading of two masks for every image due to its aggregation over the \texttt{CP} values of the masks. 
For Q2, Q4, and Q5 on \textit{ImageNet}, TileDB is slower than the other two baselines. 
The reason is that TileDB has to sequentially load masks from the disk (instead of slicing the same ROI from multiple masks at once) because the ROIs in these queries are mask-specific. 
This results in suboptimal disk read bandwidth utilization. 
During the execution of all queries on PostgreSQL and NumPy and for the other queries on TileDB, we observed that the disk read bandwidth was fully utilized, reaching 125 MiB/s, the provisioned disk read bandwidth for our EBS volumes. 
This confirms that the query execution time is dominated by the time required to load the masks from disk. 
Therefore, any system that does not reduce the number of masks loaded from disk during execution can achieve, at best, a comparable query time to that of NumPy and PostgreSQL. 
And, while faster EBS volumes could enhance the baselines' performance, \system would still outperform them by reducing mask-loading during query execution.

\system executes each query in under 5 seconds on \textit{WILDS} and in less than 20 seconds on \textit{ImageNet}, providing query time speedups of up to two orders of magnitude over the baselines. 
This significant difference in performance is attributed to \system loading many fewer masks (shown in \cref{tab:masks-loaded}) because its filter-verification framework enables it to avoid loading from disk the masks that are guaranteed to satisfy the query predicate or guaranteed to fail it. 
On \textit{ImageNet}, \system's query time for Q4 is longer compared to other queries, even though the number of masks loaded for Q4 is smaller. 
This discrepancy stems from the additional computation \system performs for Q4 ($2\times$ bound computation than other queries), as it contains an aggregation.

\masksLoadedTable

\subsection{Performance on Different Query Types} \label{sec:eval-single-query-workload-types}
\queryTimeVsQueryTypeFigure

In this experiment, we evaluate the performance of \system on three types of queries with varying parameters.
We only show the execution times of \system because, for each query type, baseline methods have similar execution times as the queries of the same type in \cref{sec:eval-single-query-workload-motivation}, regardless of specific query parameters.
For each dataset and query type, we generate $500$ queries with randomized parameters and execute them using \system:

\begin{itemize}[itemsep=1pt, topsep=1pt, leftmargin=10pt]
    \item \textbf{Filter}: this query type contains mask selection queries with a filter predicate $\texttt{CP}(mask, roi, (lv, uv)) > T$. For every query, $roi$ is set as the foreground object bounding box in a mask generated by YOLOv5 \cite{yolov5}. $lv$ and $uv$ are randomly selected from $[0.1, ..., 0.9]$ and $uv$ is always greater than $lv$. The count threshold $T$ is randomly chosen from $[0, 1, ..., \text{total \# pixels}]$.
    \item \textbf{Top-K}: this query type returns masks ranked by $\texttt{CP}(mask, roi, (lv, uv))$. For each query, $roi$ is randomly generated as any rectangle within the masks. This $roi$ is generated once for each query and remains constant across all masks. $k$ is set to $25$. The order of query result, i.e., \texttt{ORDER BY ... DESC} or \texttt{ASC}, is randomly selected for each query.
    \item \textbf{Aggregation}: this type of query returns images ranked by $\texttt{mean}(\texttt{CP}(mask, roi, (lv, uv)))$ of multiple masks associated with each image. Two masks are associated with each image and they are generated by GradCAM based on different models. $k$ is set to $25$. $roi$, $lv$, $uv$, and the order of the query result is randomly selected for each query.
\end{itemize}

\queryTimeVsFractionOfMasksLoadedFigure

\combinedBoundSegmentsFigure

\cref{fig:query-time-vs-query-type} shows the distribution of query execution times for each query type on both \textit{WILDS} and \textit{ImageNet}.
The figure displays the median, minimum, maximum, and interquartile range (IQR) of these times.
The whiskers represent outliers, which are defined as values that are more than 1.5 times the IQR away from the median.

\system demonstrates its superior query execution performance across all query types with varying parameters.
Even when considering the worst-case execution time (i.e., the outliers), \system would still outperform the baselines by a considerable margin, because the baselines would still load all masks from disk and process them, regardless of the query parameters.

Moreover, we find that the query execution times of \system do not exhibit a strong correlation across different query types.
We note that the 75th percentile of the \textit{Filter} query type has a longer execution time than that of the other two query types.
This is because, for the other query types, \system compares the bounds (of \texttt{CP}) with the \texttt{CP} values of the current top-$k$ set (k=25). 
This process generally allows for more efficient mask filtering than comparing the bounds with a fixed count threshold $T$ in the \textit{Filter} query type. 
For example, on \textit{WILDS}, at the 75th percentile in query time, the number of masks pruned in \system's filter stage during query execution is \num{21184} for the \textit{Filter} query type, \num{22106} for \textit{Top-K}, \num{21677} for \textit{Aggregation}. 

Instead, we observe that the execution times tend to differ more significantly among queries with different parameters within the same query type.
In fact, as we discuss further in \cref{sec:eval-query-time-analysis}, for a given dataset, the query execution time of \system is primarily determined by the fraction of masks loaded (FML), i.e., masks that are loaded from disk and used to compute its \texttt{CP} value during query execution.
The difference in execution times within the same query type is mainly due to the difference in the FML for each query.
For example, for the \textit{Filter} query type on \textit{WILDS}, the FML at the 25th, 50th, and 75th percentiles are $0.002$, $0.012$, and $0.049$, respectively.

\vspace{-1.0em}
\subsection{\system's Query Time Analysis} \label{sec:eval-query-time-analysis}

In this section, we explore factors affecting \system's query execution time by analyzing \num{1500} \textit{Filter} queries, defined in \cref{sec:eval-single-query-workload-types}, executed by \system on each dataset.

With \cref{fig:query-time-vs-fraction-of-masks-loaded}, we first establish that, given a dataset, \system's query execution time is proportional to the fraction of masks loaded (FML) for each query. 
The FML for a query is defined as the ratio of masks loaded from disk and used to compute their actual \texttt{CP} values to the total number of masks in a dataset. 
The Pearson's correlation coefficient between query time and FML is $0.99$ for \textit{WILDS} and $0.96$ for \textit{ImageNet}. 
It again corroborates that query execution time is dominated by loading masks from disk and computing their \texttt{CP} values, with a higher FML indicating more masks being loaded from disk.

Now that we have established the relationship between query execution time and FML, we investigate the factors that affect FML, including the query parameters (region of interest $roi$, pixel value range $(lv, uv)$, count threshold $T$), data in the masks ($mask$), and index granularity (index size).
For \system, FML is the fraction of masks that are neither pruned nor added directly to the result set by the filter stage in the filter-verification framework.
FML corresponds to \textit{Case 3} in Step 2 of the filter stage; for each mask belonging to this case, its lower bound $\ubar{\theta}$ for \texttt{CP} computed by \system is not greater than the count threshold $T$ and its upper bound $\bar{\theta}$ for \texttt{CP} is greater than $T$, i.e., $\underline{\theta} \leq T < \bar{\theta}$.

\cref{fig:combined-bound-segments} shows the distribution of bounds computed by \system for both datasets and queries with varying parameters from the \num{1500} \textit{Filter} queries analyzed.
Each subfigure shows the distribution of bounds for a different (dataset, index size, $(lv, uv)$) combination.
The $roi$ for all subfigures is the foreground object bounding box.
The  (vertical) segments in each subfigure represent the bounds computed by \system for \num{1000} masks randomly sampled from the dataset.
Each red horizontal dashed line represents an example count threshold $T$.
In this way, each subfigure visualizes the relationship between the bounds and FML: for each count threshold $T$, FML equals the fraction of the segments intersecting with the red dashed line defined by $T$.

In each subfigure, different count thresholds $T$ lead to varying FMLs for the same dataset, index size, and query parameters, as the fraction of segments intersecting with the red dashed line changes.

Comparing subfigures with the same $roi$ and $(lv, uv)$ but on different datasets reveals that different sets of masks can result in different FMLs for the same query parameters because of different pixel value distributions in the $roi$ of the masks.
Similarly, changing $roi$ essentially alters the set of masks targeted by the query, leading to different FMLs.
Subfigures with the same dataset and $roi$ but different $(lv, uv)$ configurations also exhibit different bound distributions and FMLs for the same count threshold $T$.

Moreover, subfigures sharing the same dataset and $(lv, uv)$ but with varying index sizes display different bound distributions and FMLs as well.
Larger index sizes offer more granular indexes, tighter bounds (shorter vertical segments in the figure), and lower FMLs for the same query parameters.
For example, comparing \cref{fig:combined-bound-segments} (a) and (c), we observe that the bounds computed by \system for \textit{WILDS} with $(lv, uv) = (0.6, 1.0)$ are tighter for the larger index size.
Therefore, the FML for the same count threshold $T$ is lower for the larger index size.

In conclusion, the data in the masks, region of interest $roi$, pixel value range $(lv, uv)$, and index size determine the distribution of bounds computed by \system. The count threshold $T$ defines the FML given the distribution of bounds, and the FML dictates the query execution time of \system.
The granularity of the index represents a trade-off between index size and query time, depending on user application requirements and available resources.

\vspace{-1em}
\subsection{Multi-Query Workload Performance} \label{sec:eval-multi-query-workload}

In this section, we evaluate \system on multi-query workloads with and without the incremental indexing technique (\cref{sec:incremental-indexing}) which mitigates \system's potential start-up overheads.
We generate workloads to simulate the exploration and analysis processes of users who seek to identify sets of masks satisfying a given predicate.

We simulate workloads where a user begins with a query targeting masks of image subsets belonging to certain classes and then progressively explores masks associated with other classes.
For example, to identify images with spurious correlations (\cref{sec:example-queries}), the user may first look at the confusion matrix and identify classes with high false positive rates.
Then, the user may issue queries to retrieve images predicted as those classes to identify possible spurious correlations.
Several queries may be issued targeting those masks, as different query parameters (e.g., $roi$, $lv$, $uv$, $T$) may be used to retrieve and rank masks with different properties, e.g., masks focusing on the foreground object and masks focusing on the background.
After analyzing the returned masks, the user may continue to explore masks of other classes and repeat the process.

\multiQueryWorkloadFigure

To account for this behavior, we generate four different workloads for each dataset, each of which comprises 200 \textit{Filter} queries, with query parameters randomly generated following the approach described in \cref{sec:eval-single-query-workload-types}.
A parameter $p_{seen}$ is associated with each workload, representing the likelihood of querying previously targeted masks within the same workload.
Randomized query parameters and $p_{seen}$ are intended to simulate the user's behavior of issuing multiple queries targeting the same set of masks with different parameters to retrieve masks having different properties.
Additionally, each query within a workload targets a specific subset of masks (e.g., masks of images predicted as certain classes) from the corresponding full dataset.
Let $N$ denote the total number of masks within a dataset.
The number of masks targeted by each query, $n$, is randomly chosen from $[0.1 \cdot N, 0.2 \cdot N, 0.3 \cdot N]$.
Then, the set of targeted masks is generated as follows, we sample without replacement $n$ masks consisting of $p_{seen}\%$ targeted masks and $(1 - p_{seen})\%$ unseen ones.
Note that when the number of remaining unseen masks is less than $n \cdot (1 - p_{seen})$, we include all the unseen masks in the current query and switch to only sampling seen masks for the remaining queries in the workload.

The workloads are labeled as Workload 1, 2, 3, and 4, with their respective $p_{seen}$ values set to $0.2$, $0.5$, $0.8$, and $1.0$.
These probabilities signify varying levels of dataset exploration, with Workload 1 exhibiting the highest degree of exploration and Workload 4 exhibiting the lowest.
By evaluating \system's performance across these diverse workloads, we aim to assess its effectiveness under a range of dataset exploration scenarios.

\cref{fig:multi-query-workload} shows the performance of \system on these four workloads for both \textit{WILDS} and \textit{ImageNet}.
\system is evaluated with and without incremental indexing against NumPy which represents existing methods that must load and process all masks from disk for each query.
In the figure, MS-II refers to \system with incremental indexing and MS refers to \system without incremental indexing.
We measure the cumulative total time, i.e., the time elapsed for index building plus the time elapsed for query execution, for each method.
\cref{fig:multi-query-workload} shows the result.
Note that the time to initially build the indexes without incremental indexing is included with the $0$-th query for MS in all subfigures.

\cref{fig:multi-query-workload} (a) and (b) show the cumulative total times for Workload 2.
The results for other workloads are not shown because MS and NumPy have similar performance trends across all workloads.
MS exhibits a slow growth in cumulative total time because it executes all queries efficiently with the filter-verification query processing framework.
However, it incurs a start-up overhead due to the need to build indexes for all masks in the dataset ahead of time.
In contrast, NumPy has no start-up overhead but suffers from rapid growth in its cumulative time because it does not reduce the required work for each query.
Nevertheless, the cost of building the indexes for MS is quickly amortized across the queries thanks to the filter-verification query processing framework and the CHI technique.
On both datasets, MS outperforms NumPy after approximately 10 queries.
MS-II strikes a good balance between MS and NumPy, eliminating the start-up overhead while achieving comparable query execution times to MS.

\cref{fig:multi-query-workload} (c) and (d) show the ratio of cumulative total time between MS-II and MS for all workloads on both datasets.
We first discuss the results for Workload 1, 2, and 3.
For both datasets, we observe that this ratio grows rapidly at the beginning for Workload 1, 2, and 3, and then peaks at around 10 to 20 queries before decreasing gradually.
The initial fast growth is due to the fact that for the first few queries, MS-II needs to answer them without the help of indexes for the unseen masks targeted, which is similar to the behavior of NumPy, and to build indexes for these masks.
Among workloads, Workload 1 has the highest growth rate in this ratio because it has the lowest $p_{seen}$ value, resulting in more unseen masks being targeted during the first few queries and therefore forcing MS-II to build indexes for more masks.
Then, the ratio peaks at around 10 to 20 queries because, at this point, MS-II has built indexes for all the masks in the dataset, and subsequent queries can be executed using the filter-verification framework without index building.
The peak ratio exceeds 1.0 because MS-II must load the masks from disk and compute their \texttt{CP} values during query execution the first time they are targeted.
In contrast, MS utilizes pre-built indexes for all targeted masks in all queries, which results in a lower cumulative total time.
Then, after the peak, the ratio decreases gradually because the cumulative total time for MS-II grows at a similar rate to MS's cumulative total time.

For Workload 4, on both datasets, MS-II never completes building the indexes for all masks, as only 30\% of the masks in the dataset (\num{6683} for \textit{WILDS} and \num{399351} for \textit{ImageNet}) are eventually targeted by all the queries in this workload.
As a result, after the rapid initial growth, the ratio of cumulative total time plateaus.
This ratio never reaches 1.0 because the time spent by MS to build the indexes for the never-targeted masks is not amortized across queries.

Lastly, we note that users typically pause between queries to examine results.
Hence, \system can leverage this interval to compute indexes, yielding better user-perceived latencies. 

\section{Related Work}
\label{sec:related-work}

\textbf{Image masks in ML tasks.}
Masks are widely used in ML to annotate image content, e.g., saliency maps~\cite{sundararajan2017axiomatic, smilkov2017smoothgrad, gradcam2017, zhou2015cnnlocalization} and segmentation maps~\cite{he2018mask, kirillov2023segment, ronneberger2015unet}.
Practitioners use them for a variety of applications, including identifying maliciously attacked examples~\cite{ye2020detection, wang2022adversarial, zhang2018detecting}, detecting out-of-distribution examples~\cite{hornauer2022heatmapbased}, monitoring model errors~\cite{meerkat2023goel, kangdata, tesla2020gritti}, and performing traffic and retail analytics~\cite{datafromsky-traffic, datafromsky-retail}.
These applications motivate the design of \system and could utilize \system's efficient query execution to quickly retrieve examples that satisfy the desired properties.

\noindent\textbf{Data systems for ML workloads and queries.}
Numerous systems have been proposed to better support ML workloads and queries~\cite{boehm2019systemds, asada2022share, miao2017modelhub, vartak2016modeldb, xin2018accelerating, he2022query, phani2021lima, gharibi2019automated, masq2021}.
\system is related to systems that support the inspection, explanation, and debugging of ML models~\cite{sellam2019deepbase, wu2020complaint, mehta2020toward, vartak2018mistique, deepeverest2021he}.
Among these, DeepEverest~\cite{deepeverest2021he} is the closest to \system.
It is designed to support the efficient retrieval of examples based on neural representations, helping users better understand neural network behavior.
While \system also focuses on efficiently retrieving examples, it targets queries based on mask properties rather than neural representations.

\noindent\textbf{Image databases and querying.}
Many systems and techniques support efficient queries over image databases~\cite{beaver2010finding, vdms2021remis, qbic1995flickner, bhute2014content, ii2001survey}. 
However, these methods are not optimized for our target queries. 
For example, VDMS~\cite{vdms2021remis} focuses on retrieving images based on metadata, while DeepLake~\cite{hambardzumyan2022deep} supports content-based queries but lacks support for querying based on aggregations over pixels. 
Array databases like SciDB~\cite{scidb2010brown} are designed for handling multi-dimensional dense arrays but do not efficiently support searching through large numbers of arrays. 
In contrast to \system, these existing systems do not reduce the work required to execute our target queries. 
Moreover, existing multi-dimensional indexes, discussed in \cref{sec:challenges}, are ill-suited for dense data like masks and fail to accommodate mask-specific regions of interest in queries.

\vspace{-1.0ex}
\section{Conclusion}
\label{sec:conclusion}

We introduced \system, a system that accelerates queries that retrieve examples based on mask properties. 
By leveraging a novel indexing technique and an efficient filter-verification execution framework, 
\system significantly reduces the masks that must be loaded from disk during query execution. 
With around $5\%$ of the size of the dataset, \system accelerates individual queries by two orders of magnitude and consistently outperforms existing methods on various multi-query workloads. 

\end{sloppypar}

\pagebreak
\balance
\bibliographystyle{ACM-Reference-Format}
\bibliography{paper}

\end{document}